\documentclass[times,twocolumn,final]{elsarticle}

\usepackage{framed,multirow}

\usepackage{amssymb}
\usepackage{latexsym}
\usepackage{amsmath}
\usepackage{url}
\usepackage{xcolor}
\definecolor{newcolor}{rgb}{.8,.349,.1}

\usepackage{comment}  

\usepackage{hyperref}
\usepackage[nameinlink,noabbrev,capitalize]{cleveref}
\usepackage{xr}

\usepackage[switch,pagewise]{lineno} 

\journal{Combustion and Flame}

\begin{document}


\begin{frontmatter}

\title{Insight into the combustion dynamics of ITA-driven swirl flames}%

\author[1]{SK Thirumalaikumaran\fnref{fn1}}
\emailauthor{sbasu@iisc.ac.in}{Corresponding Author Saptarshi Basu}
    
\author[1]{Balasundaram Mohan\fnref{fn1}}
\author[1,2]{Saptarshi Basu}
\fntext[fn1]{These authors contributed equally to this work.}  

\address[1]{Department of Mechanical Engineering, Indian Institute of Science Bangalore, India.}
\address[2]{Interdisciplinary Centre for Energy Research, Indian Institute of Science Bangalore, India.}

\begin{abstract}
In this article, we examine the flame, flow, and acoustic coupling of Intrinsic ThermoAcoustic (ITA) driven combustion instability using a high-shear swirl injector in a model combustor. The combustor is operated using pure methane and methane-hydrogen mixture as fuels with air in a partially premixed mode. In this study, the airflow rate is varied by keeping the fuel flow rates constant, corresponding to the Reynolds number range of 8000-19000 and a fixed thermal power of 16 kW. Acoustic pressure in the combustion chamber, high-speed OH$^*$, CH$^*$ chemiluminescence images, high-speed particle image velocimetry, and steady exhaust gas temperature are measured for scrutiny. The combustor shows non-monotonic variations in the acoustic pressure amplitude for both fuels. A wider operating envelope and relatively large amplitude acoustic fluctuations are observed for the methane-hydrogen mixture compared with pure methane. Both methane and methane-hydrogen mixtures exhibit a linear increase in dominant instability frequencies with airflow, indicating the ITA-driven combustion instability. A low-order network model is developed that incorporates a simple $n-\tau$ flame response. It reproduces the observed dominant instability frequency by taking in the convective time delay from the experiments that substantiates the ITA-driven instability. Spatio-temporally resolved flame and flow dynamics show periodic axisymmetric vortex shedding from the outer shear layer and its subsequent interaction with CRZ during large amplitude acoustic oscillations. At relatively low amplitude, the vortices shed from both inner and outer shear layers convect downstream at different velocities without interacting with each other. The periodic vortex shedding process across the operating condition is further examined using a simplified model representing vortex dynamics. The findings of this study shed light on the ITA-driven combustion instability and the influence of azimuthal air staging on the dynamic coupling relevant to modern gas turbine injectors.
\end{abstract}


\end{frontmatter}


\section*{Novelty and significance statement}
The novelty of this study lies in providing detailed insight into the flow and flame dynamics of the ITA-driven combustion instability for the methane-hydrogen mixture for the first time. In this context, we explored both methane and methane-hydrogen combustion dynamics in a model combustor. Our findings provide valuable insight into the coupling mechanisms between heat release rate fluctuations and vortex dynamics in annular swirl injectors, which are integral to modern gas turbine systems. These insights presented here form a foundational step toward the development of passive control strategies for ITA-driven combustion instability during the clean energy transition.

\section{Introduction \label{sec:Intro}}
 It is well known that the combustion of clean fuels are considered in the transportation and aviation industries. Further, the contemporary aspects of clean energy transition, sustainable combustion, hydrogen-blending with natural gas for reduced NO\textsubscript{x} emission, and extended operation envelope require the understanding of natural gas-hydrogen combustion. The hydrogen fuel-air mixture is vulnerable to auto-ignition, flashback, lean blow-out, high adiabatic flame temperature, increased NO\textsubscript{x} emission and susceptible to combustion instabilities \cite{beita2021thermoacoustic}. To circumvent these issues, the researchers developed numerous fuel injection strategies resulting in various swirl injector designs \cite{chterev2021effect,marragou2022stabilization,katoch2022dual} and also possible consideration of porous media burners \cite{lee2022high} and micromix burners \cite{funke202130}. This study is motivated by these recent developments in injector design and characterization for aero-derivative and industrial gas turbine combustors considering hydrogen and natural gas. Specifically, the present study compares the combustion dynamics of methane and methane-hydrogen blending in a high shear swirl injector employed in model combustor encountering ITA mode. The combustion instability manifests in the form of fluctuations in flow properties and heat release rate imposing cyclic mechanical and thermal load on combustion chamber walls. They cause potential premature components or engine failure, resulting in costly shutdowns and maintenance scenarios \cite{lieuwen2005combustion}. In the following paragraphs, we summarize the studies relevant to hydrogen addition to conventional fuels or pure hydrogen combustion.

Schefer et al. \cite{schefer2002combustion, schefer2003hydrogen} studied lean premixed swirl-stabilized flames with hydrogen-natural gas blends. They observed increase in OH radicals near the Lean Blow-Out (LBO) limit, reduced CO emissions, and unchanged NO\textsubscript{x} emissions at longer residence times. The addition of hydrogen enhanced the flammability limits, altered the flame structure, and increased the flame blowout velocity for stoichiometric mixtures. Figura et al. \cite{figura2007effects} studied lean premixed methane-hydrogen flames under varying fuel mixtures, inlet temperature, inlet velocity, combustor length, and equivalence ratio using center of heat release. The results revealed a unified path for stable and unstable operating conditions. Hydrogen addition extended stability toward leaner conditions with reduced acoustic oscillations. Kim et al. \cite{kim2009hydrogen} investigated swirl and hydrogen effects on confined flames, finding improved lean stability and increased NO\textsubscript{x} due to smaller recirculation zones. However, increased swirl or excess air mitigated NO\textsubscript{x} emissions by enhancing cold air entrainment and lowering flame temperature. Emadi et al. \cite{emadi2012flame} observed a reduction in flame surface curvature as hydrogen was added, attributed to a wrinkled flame surface in a low swirl premixed combustor. In addition, the authors reported an increase in the flame surface density, which was amplified at high pressures. Lee et al. \cite{lee2015investigation} studied hydrogen, CO, and methane-air mixtures in a swirl combustor and observed shorter flame spread to the Outer Recirculation Zone (ORZ) and higher frequency oscillations in hydrogen-rich cases. The authors explained the observed multi-mode instability frequency using the centroid of flame length and skewed time delay.

Karlis et al. \cite{karlis2019h2} showed that adding over 10\% hydrogen to natural gas in a lean premixed swirl combustor triggered thermoacoustic oscillations. They used extinction and flow-induced strain rates to delineate thermoacoustic instability at given equivalence and Reynolds numbers. Beita et al. \cite{beita2025effect} investigated hydrogen addition up to 50\% by volume in a dual-can combustor and observed a transition from stable to self-excited thermoacoustic instability. During limit cycle oscillations, strong out-of-phase and weak in-phase multimodal interactions between the cans were identified. Chterev and Boxx \cite{chterev2021effect} found that hydrogen enrichment up to 50\% transforms flame shape from M to V through the bistable flame. They showed that initial phase delay determines whether hydrogen enrichment leads to stable or unstable combustion. Ji et al. \cite{ji2024structure} demonstrated that higher equivalence ratios, swirl numbers, and hydrogen enrichment (up to 80\%) enhance combustion instability. With increasing hydrogen addition, the flame transitions to an M-shape, showing greater stabilization in the ORZ, along with increased flame wrinkling and surface area, as revealed by OH-PLIF measurements. Agostinelli et al. \cite{agostinelli2023large} found that hydrogen addition at atmospheric pressure induces combustion instability and M to V flame transition. At elevated pressure, the flame become compact, thermoacoustically stable with a lifted M to attached M flame transition. Taamallah et al. \cite{taamallah2015thermo} studied the relation between flame macrostructures and thermoacoustic modes using methane–hydrogen mixtures under coupled and decoupled conditions. Four distinct flame shapes were identified, each corresponding to specific acoustic modes. Hydrogen addition up to 20\% shifts these transitions to lower equivalence ratios and widens the stability window. Thirumalaikumaran et al. \cite{Thiru2025_ECM} studied the effect of flame macrostructures and thermoacoustic instability using a high shear injector by varying the flare angles. The phase-averaged Rayleigh index is applied to identify regions that contribute to the thermoacoustic driving. At a flare angle of $\beta=40^\circ$, the Rayleigh index indicates that jets and the inner shear layer are the primary instability drivers causing high-amplitude acoustic oscillations. In contrast, the $\beta=90^\circ$ shows lower amplitude oscillations driven by the Inner Recirculation Zone (IRZ) and the wake.

Cegile et al. \cite{ceglie2023thermoacoustic} used CFD and FEM to analyze flow field and flame dynamics in bluff-body stabilized combustors. They observed reduced IRZ size, time delay, and higher heat release rates for hydrogen, linked to enhanced instability growth. Garcia et al. \cite{garcia2024impact} computationally examined hydrogen-enriched partially premixed flames. They showed that velocity and equivalence ratio fluctuations interact during mixing process leading to equivalence ratio fluctuations, causing overall heat release rate fluctuations. Ann et al. \cite{ahn2024longitudinal} experimentally investigated a hydrogen-methane mixture in a pressurized annular combustor, examining thermoacoustic response and the transition between longitudinal and azimuthal modes under steady and transient conditions using time series analysis. Anestad et al. \cite{aanestad2024mitigating} introduced secondary injection to reduce primary flow velocity, altering flame structure and combustion dynamics. Secondary combustion either constructively or destructively couples with the primary combustion leading to changes in the combustion dynamics. Link et al. \cite{link2025experimental} studied discrete fuel jets in a swirl cross-flow along with axial air injection for dual fuel capability, transitioning from methane to hydrogen while maintaining constant thermal power. They observed changes in flame structure and flow field when the non-swirling axial air jet was injected. Depending on the amount of axial air injection the authors were able to enhance the blowout and flashback limits.

Katoch et al. \cite{katoch2022dual} investigated radial fuel inhomogeneities in a dual-swirl burner with hydrogen-ammonia mixtures. The radial inhomogeneity variations altered the swirl number, flame shape, and thermoacoustic behaviour. Least fuel inhomogeneity led to relatively strong thermoacoustic signatures associated with ITA modes and localized heat release under high inhomogeneity. Kumar et al. \cite{kumar2025intrinsic} showed that hydrogen addition enhances interaction index and reduces time lag, promoting ITA modes and exhibit a change in axial pressure gradient across the flame. Their 1D model shows that modifying geometry, particularly the injection tube area, can stabilize ITA modes. Ghani et al. \cite{ghani2021control} investigated methane-hydrogen mixtures with up to $50\%$ hydrogen to control ITA modes using DNS and network model. Hydrogen addition shortens flame length, increases ITA frequencies, reduces Flame Transfer Function (FTF) gain and stabilizing the flame at $50\%$ H$_2$ content. This points out the existence of thermoacoustic instability with duct acoustic modes were excited to large amplitude oscillations. It is clear that only a limited number of studies reported the influence of hydrogen addition on the ITA modes in practical combustor configurations in the past mainly involving computation and low order model lacking rigorous experimental campaign using clean fuels. Furthermore, it is not clear from the literature how the swirl flow field is affected under the hydrogen addition for ITA driven modes. Therefore, in this study, we examine in detail the flame and flow field dynamics of ITA mode for methane and methane-hydrogen fuel-air mixtures. For this purpose, a high shear counter rotating swirler with a fixed thermal power of 16 kW operated in a partially premixed mode. The results show the convective scaling of dominant instability frequency for both methane and methane-hydrogen mixture.

The article is structured as follows. In Section \ref{sec:Exp}, we describe the experimental setup, data acquisition, and high-speed diagnostics techniques. 
A low-order model is introduced in Section \ref{sec_net_model} to substantiate the presence of the ITA mode. Dynamic Mode Decomposition (DMD) analyses of the flame and flow fields are discussed in Sections \ref{sec_Flm_DMD} and \ref{sec_PIV_DMD}, respectively. The influence of outer and inner swirl jets on vorticity dynamics is examined in Section \ref{vor_dyn}. Finally, key findings of this study are summarized in Section \ref{con}.

\section{Model combustor and diagnostics facilities \label{sec:Exp}}
In this section, experimental setup, data acquisition system, high-speed PIV, high-speed chemiluminescence imaging, and data post-processing are described.

\subsection{Experimental setup}
\begin{figure}[h!]
    \centering
    \hspace*{-3 mm}
    \includegraphics[width=0.51\textwidth]{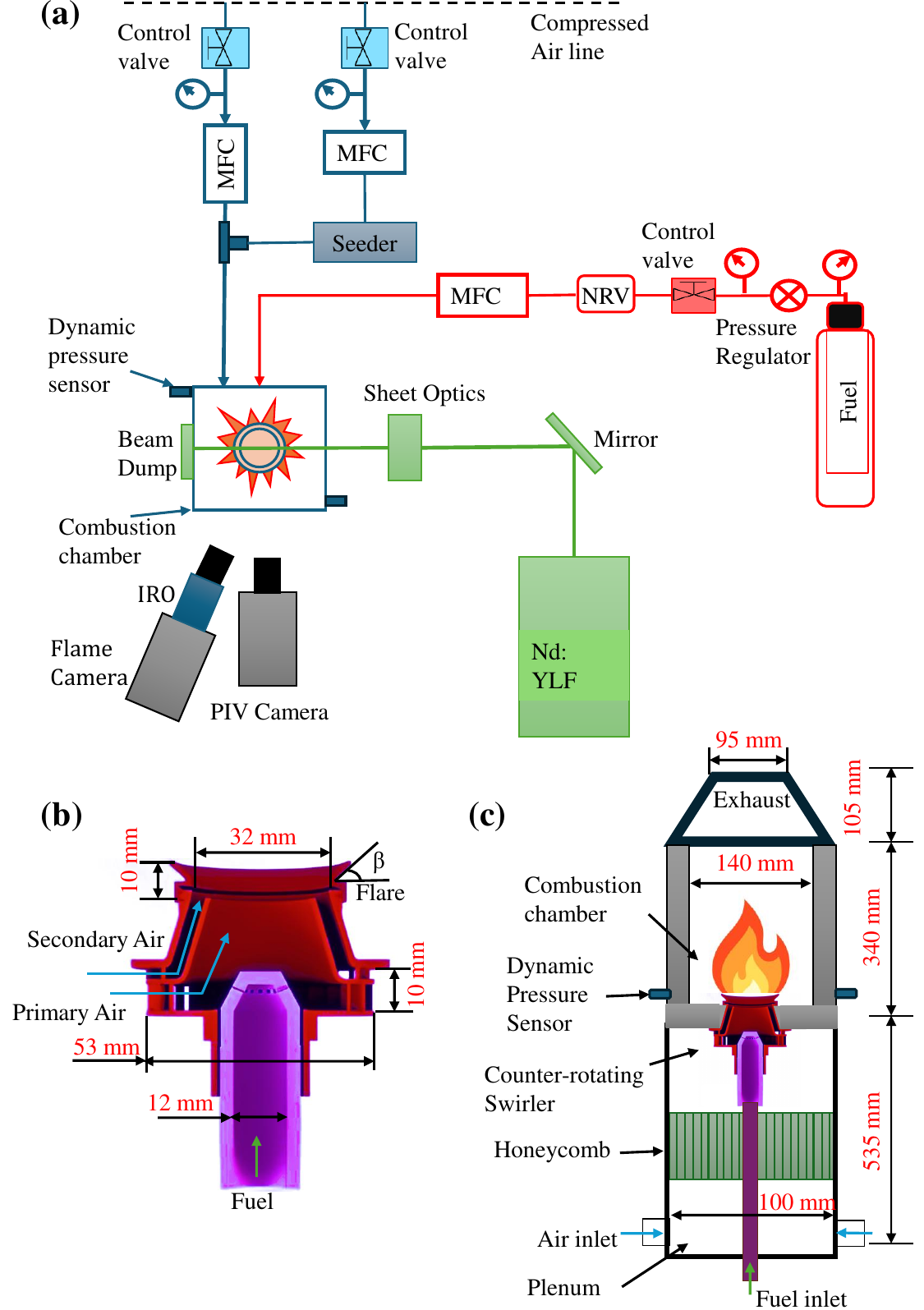}
    \caption{(a) Schematic of the flow lines and high-speed optical diagnostics arrangements. The combustion chamber is shown by a square cross-section with dynamic pressure sensors mounted near the dump plane at opposite locations. The high-speed cameras used for flame imaging and PIV alignments are shown for clarity. The air (along with alumina seeder) and fuel flow lines are depicted using blue and red colors respectively. (b) Shows the cut section of a high-shear counter-rotating radial swirler with a central fuel lance. (c) Schematic representation of the combustor arrangement with dimensions. The combustor operates in a partially premixed mode. The location of fuel injection to the flare tip ($l_s$) is 30 mm.}
    \label{exp_line}
    \vspace{-2mm}
\end{figure}

The experimental studies are carried out in a model combustor consisting of a plenum, fuel lance, radial swirler, combustion chamber, and exhaust nozzle depicted in Fig. \ref{exp_line} (a-c). Air is supplied from a high-pressure storage tank and regulated using an Alicat mass flow controller having a range of 0–500 slpm. The air then flows through a 100 mm diameter plenum ($d_p$) equipped with a honeycomb structure to ensure uniform flow prior to reaching the swirler. Another Alicat mass flow controller independently controls the fuel flow rate which has a range of 0–50 slpm. Methane (CH$_4$) and methane-hydrogen (CH$_4$-H$_2$) mixture (on a $50\%:50\%$ volumetric basis) are used as fuels in this experimental study. These fuels are injected through a centrally located fuel lance (refer Fig. \ref{exp_line} (b,c) for clarity). These air and fuel flow lines are shown in Fig. \ref{exp_line} (a) with blue and red colours.

The combustion chamber, having a square cross-section of 140 mm × 140 mm with a length ($l_c$) of 340 mm, features UV-grade fused silica windows and a quartz plate to manage thermal loads. This further facilitates the flow and chemiluminescence imaging. Exhaust gases pass through a 95 mm × 95 mm linearly converging nozzle having a length ($l_{n}$) of 105 mm. The nozzle is connected at the combustion chamber exit to avoid backflow through entrainment from the surroundings. The nozzle flow is not choked in this study. The summary of major combustor dimensions and important flow properties are depicted in Fig. \ref{exp_line}.

The flame is stabilized inside the combustion chamber using a swirl injector in a partially premixed mode. The cut section of the injector comprising of the swirler and central fuel injection tube, is provided in Fig. \ref{exp_line} (b). The air and fuel mixing occurred in the primary region before exiting through a diverging flare section having a diameter ($d_f$) of 32 mm. In this study, the divergence (or flare) angle of $\beta=60^\circ$ is considered and the combustion studies using other flare angles for pure methane is reported in \cite{Thiru2025_ECM}. The airflow enters the counter-rotating swirlers with a 6:4 split ratio, where 60\% of air passed through the primary vanes and 40\% air through the secondary vanes. The geometric swirl numbers of the primary and secondary swirlers are 1.5 and 0.78, respectively.

The swirler assembly accommodates a fuel injection nozzle as shown in Fig. \ref{exp_line} (b). The fuel exits as 12 radial jets having 0.6 mm diameter, arranged concentrically and oriented at 55° vertically to the nozzle axis. Experiments were conducted under atmospheric pressure conditions, with Reynolds numbers ranging from 8000 to 19000 with a fixed thermal power of 16 kW for both methane and methane-hydrogen mixtures. The Reynolds number is calculated as,
\begin{equation}
Re = \frac{\rho V d_s}{\mu}
\end{equation}
where $d_s$ is the swirler diameter. The exit velocity $V$ of the swirler is calculated from the inlet mass flow rate of air.

\subsection{High-speed Particle Image Velocimetry (PIV)}
A high-speed Particle Image Velocimetry (PIV) system is employed to capture the flow field within the combustion chamber. Imaging was performed using a high-speed CMOS camera (SA-5, 7000 frames per second, 1024 × 1024 pixels) from LaVision. For this study, the camera is operated at a frame rate of 3000 Hz and a resolution of 1024 × 1024 pixels. The camera is equipped with an AT-X M100 Tokina f/2.8 lens, and 527$\pm$10 nm bandpass filter to optimize imaging clarity and accuracy. The flow was seeded with micron-sized alumina particles ($\sim$1 µm) having relaxation time of 1.2$\times 10^{-5}$ s and maximum velocity error caused by particle slip is of 0.4\% to facilitate flow visualization. A high-speed dual head Nd: YLF laser (DM30 series) with 30 mJ per pulse energy, 527 nm wavelength, and 10 kHz illuminates these seeder particles, enabling their capture by high-speed camera. The laser beam is transformed into a thin sheet approximately 1 mm thick using a combination of telescopic lens and a cylindrical lens within the sheet optics system. The PIV camera along with laser system is depicted in Fig. \ref{exp_line}(a) with green colour line. Flow field data were post-processed using DaVis 8.4 commercial software to generate velocity vector fields. The processing method employed a multi-pass cross-correlation approach. An initial interrogation window of 64 × 64 pixels with 50\% overlap was used, followed by a final interrogation window of 32 × 32 pixels with 75\% overlap.

\subsection{High-speed chemiluminescence flame imaging}
A LaVision high-speed SA-5 CMOS camera with High-Speed Intensified Relay Optics (HS-IRO) was used to capture OH$^*$ chemiluminescence from the flame. The system utilized a UV-sensitive lens (Cerco, 105 mm, f/2.8) paired with a 310 $\pm$ 10 nm bandpass filter to isolate the desired spectral range. The HS-IRO was triggered via a LaVision Programmable Timing Unit and synchronized with the PIV imaging system. High-speed images are recorded at a resolution and frame rate of 1024 x 1024 pixels and 3000 Hz for 1 second. Commercial Davis software and MATLAB software are used to post-process the chemiluminescence images. In addition, to calculate the convective time delay used in section \ref{sec_net_model}, CH$^*$ chemiluminescence images triggered after high-speed PIV are acquired. For CH$^*$ imaging, the bandpass filter having $430\pm10$ nm spectral range is used. To capture the mean CH$^*$ flame images, the high speed LaVision SA-5 camera operated with a resolution and frame rate of 1024 × 1024 pixels and 60 fps, respectively. Dynamic Mode Decomposition (DMD) is applied to the OH$^*$ chemiluminescence flame images and the PIV flow field to elucidate the physical mechanism associated with the instability modes, and is discussed in the supplementary section \ref{supl:DMD}.

\subsection{Acoustic pressure and steady temperature measurements}
Two high-frequency dynamic pressure transducers (PCB 113B28) are integrated into the combustion chamber as shown in Fig. \ref{exp_line} (a, c) for acoustic pressure measurements. These pressure sensors has a sensitivity of 100 mV/Pa. These transducers are mounted on the opposite sides of the combustor, positioned 10 mm downstream of the combustion chamber dump plane. These transducers are synchronized with high-speed PIV and chemiluminescence imaging systems using the programmable timing unit. The acquired acoustic pressure data was transmitted through an NI 9205 module housed in an NI cDAQ 9179 chassis to sample at 35 kHz. 

A K-type thermocouple was employed to monitor the upstream temperature of the inlet air, positioned at 70 mm upstream of the dump plane. This thermocouple measures the temperature of the incoming air. Another K-type thermocouple is positioned at the exit of the combustor to measure the exit temperature of the product gas. Prior to data acquisition, the system was allowed to stabilize, achieving both thermal and flow equilibrium. Commercial LabVIEW software was used to acquire and visualize both the recorded acoustic pressure data and the mean inlet air temperature. Further, post-processing was carried out using MATLAB software.

\section{Results and discussion \label{sec:results}}
In this section, we examine the combustion dynamics using experimental measurements and simplified modelling in detail.

Experiments were conducted over a range of $Re$ from 8000 to 19000, corresponding to air flow rates between 200 and 450 slpm, while maintaining a constant thermal power of 16 kW and fixed fuel flow rates. The pure methane and methane–hydrogen mixture fuel flow rates are fixed and 29.5 slpm and 43 slpm, respectively. Based on acoustic pressure characterization provided in the supplementary section \ref{sec_prim_charac}, two representative cases were selected, one with higher amplitude (HA) and the other with lower amplitude (LA) for each fuel type. Furthermore, the acoustic pressure characterization in section \ref{sec_prim_charac} shows large amplitude acoustic oscillation and wider operating envelope of methane–hydrogen mixtures compared to pure methane. However, their dominant instability frequency linearly increases with air flow rate as discussed in supplementary section. In addition, the time-averaged flow field measurements from the PIV and chemiluminescence experiments revealed a smaller recirculation bubble and shorter flame height for methane–hydrogen mixture compared to the pure methane operation. Further details are provided in supplementary section \ref{avg_flame_PIV}. In the following subsection a network model is developed to examine the instability frequency associated with Fig. \ref{exp_line}(c).

\subsection{Low order network model \label{sec_net_model}}
\begin{figure}
    \centering
    \includegraphics[width=0.48\textwidth]{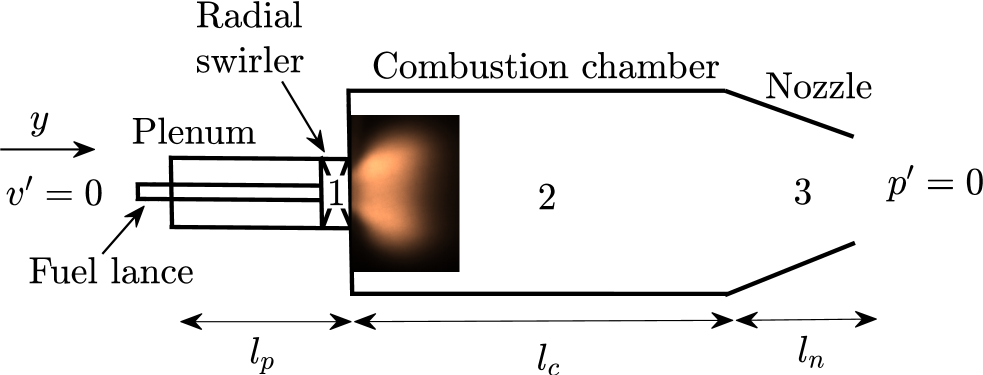}
    \caption{Schematic of the one-dimensional model combustor with various parts and dimensions marked. Fuel injection, combustion chamber, and nozzle are marked by 1, 2, and 3. Here, the lengths $l_p=535$ mm, $l_c=340$ mm, $l_n=105$ mm, and the fuel injection location to the flare tip ($l_s$) is 30 mm. The plenum diameter, swirl cup diameter, combustion chamber width and nozzle width at the exit are 100 mm, 32 mm, 140 mm, and 95 mm, respectively. The inlet $(y=0)$ and outlet $(y=L)$ boundary conditions are considered to be acoustically closed ($v^\prime=0$) and acoustically open ($p^\prime=0$) respectively. Note, $y=0$ is considered at the fuel injection location.}
    \label{fig_lom_Can_combustor}
\end{figure}

One-dimensional model combustor schematic with flame stabilized near the dump plane is shown in Fig. \ref{fig_lom_Can_combustor}. This geometry is considered for low-order network modelling discussed below with compact flame stabilized at the area change. The fundamental components of transversal modes associated with the plenum and combustion chamber are around 1 kHz and 15 kHz, respectively. Based on the experimental observations reported in Fig. \ref{fig_Prms_Fdom_SPL}(b) in supplementary section \ref{sec_prim_charac}, the dominant mode instability frequency is in the range of $97-258$ Hz. The observed values are much smaller than the transversal modes of the plenum and combustion chamber. Therefore, we assume one-dimensional plane acoustic wave propagation in the combustor along $y-$direction in Fig. \ref{fig_lom_Can_combustor}. In addition, the acoustic pressure measurements in the plenum shows no discrete peaks and hence, in the low order modeling (LOM) considered below, the plenum is not included. The LOM starts from the fuel injection location. Note that the mean flow is along the positive $y-$direction and the low-Mach number assumption is valid as discussed below \cite{dowling1995calculation}.

\begin{table}[t]
\centering
\hspace*{-3mm}
\resizebox{75mm}{!} 
{ 
\begin{tabular}{c c c c}
\hline
  $l_p=535$ mm & $l_s=30$ mm & $l_c=340$ mm & $l_n=105$ mm \\  $d_p=100$ mm & $d_f=32$ mm & $w_c=140$ mm & $w_{n,e}=95$ mm \\
  $\bar{T}_1=300$ K & $\bar{T}_2=1550 ~\&~ 1240$ K &  $\bar{T}_3=\bar{T}_2$ & n=3.5 
  \\ \hline
\end{tabular}
}
\caption{Geometrical and flow properties considered in this study remain same unless otherwise specified.}
\label{Tab1_flow_geom}
\vspace{-4mm}
\end{table}

The acoustic network consists of swirl cup where the fuel and air are injected (refer to the inset in Fig. \ref{exp_line}c), combustion chamber, and nozzle elements as shown in Fig. \ref{fig_lom_Can_combustor}. For the acoustic element with constant cross-sectional area the acoustic pressure ($p^\prime$) and acoustic velocity ($v^\prime$) are represented as follows. 

\begin{align}
p^\prime_j (y,t) & =  & i \rho_j \omega \left[A_j e^{ik_j (y-y_f)} + B_j e^{-ik_j (y-y_f)}\right] e^{-i \omega t} \\
v^\prime_j(y,t) & =  &\frac{i \omega}{ \bar{a}_j } \left[A_j e^{ik_j (y-y_f)} - B_j e^{-ik_j (y-y_f)}\right] e^{-i \omega t}
\end{align}

Here, subscript $j=1,~2$ denote swirl cup and combustion chamber, respectively, and $ i=\sqrt{-1}$. $A$ and $B$ represent the wave strength propagating in the right and left directions. Further, $k,~\omega~ (=2\pi f), ~\rho,~a,~y_f,~y$ and $t$ denotes the wave number, angular frequency, density, sound speed, flame location, mean flow direction ($0\le y \le L$) and time. The $[~]^\prime$, $\bar{[~]}$, and $\hat{[~]}$ denotes fluctuating, mean flow variables, and complex Fourier quantity, respectively. The upstream of the swirl cup is considered as an acoustically closed end.

The exhaust nozzle section has a linear reduction in width ($w_n$) along $y-$direction: $w_n=0.2986-0.4286y$ (in m). In the present study, the air is supplied with the mean temperature of 300 K corresponding to the inlet Mach number of $M_{p}<0.07$ at the swirler exit. After heat addition, the exhaust temperature measured at the nozzle exit vary between $\bar{T}_n=960-1200$ leading to the exhaust nozzle Mach number $M_3 \sim \mathcal{O}(10^{-3})$. The cross-sectional area change of the nozzle $dS_3/S_2=0.75$ and the nozzle compactness is $l_n/\lambda<0.1$. Thus, the $dS_3/S_2$ and $l_n/\lambda$ are of the same order indicating no abrupt changes in nozzle area compared with the acoustic wavelength ($\lambda$). Hence, the acoustic wave propagation inside a nozzle is considered in detail following \cite{easwaran1992plane}. Furthermore, the flow at the nozzle exit is not choked and open to the atmosphere. Hence, we assume an acoustically open boundary condition ($\hat{p}=0$) at the nozzle exit. Considering the above facts, we can safely neglect the influence of entropy wave coupling in this study \cite{marble1977acoustic,dowling1995calculation,morgans2016entropy}. In addition, the axial extent of flame is less than 77 mm which is much less than acoustic wavelength. Therefore, the flame is assumed to be an acoustically compact element and anchored at the area change.

Following the above considerations, the $p^\prime$ and $v^\prime$ are expressed as follows for a linear reduction in nozzle element \cite{easwaran1992plane}.
\begin{align}
    p^\prime_j (y,t) & =   & \frac{i \rho_j \omega}{\sqrt{S_j(y)}} \left[A_j e^{ik_j (y-y_f)} + B_j e^{-ik_j (y-y_f)}\right] e^{-i \omega t}
\end{align}
\vspace{-4mm}
\begin{multline}
   v^\prime_j(y,t) = \frac{1}{ S_j(y) } \left[ A_j (-1 + i k_j \sqrt{S(y)}) e^{ik_j (y - y_f)} \right. \\
   \left. - B_j (1 + i k_j \sqrt{S(y)}) e^{-ik_j (y - y_f)} \right] e^{-i \omega t}
\end{multline}
Here, $j=3$ and $S_3$ correspond to cross-sectional area of the nozzle calculated using $w_n$.

The acoustic flow properties across the heat source and the area change between sections $1-2$ and $2-3$ are related using the following relations \cite{dowling1995calculation}.
\begin{eqnarray}   
\hat{p}_j &  = & \hat{p}_{j-1} \\
S_j \hat{v}_{j} - S_{j-1} \hat{v}_{j-1} & = & \frac{\gamma-1}{\gamma \bar{p}} \hat{\dot{Q}}
\end{eqnarray}
Here, $\gamma,~\bar{p}$ and $\hat{\dot{Q}}$ denotes the ratio of specific heat, mean combustor pressure and Fourier amplitude of heat release rate fluctuations. Note that $\hat{\dot{Q}}=0$ in section $2-3$.

The combustion occurs in a partially premixed mode and the pressure drop across the fuel injector is about 400 kPa. Therefore we assume there is no fuel flow rate fluctuations due to combustion oscillations. On the other hand, the combustion oscillations modify the airflow fluctuations thus altering the equivalence ratio oscillations. These oscillations lead to temperature fluctuations. As we have discussed above, the combustor operates in the low-Mach number regime and the exit nozzle is not choked. Hence, the temperature fluctuations generated by the equivalence ratio fluctuations convect downstream without contributing to the observed combustion instability \cite{morgans2016entropy}. In addition, the combustor operates in the low-Mach number regime where the flame responds significantly to velocity fluctuations compared to acoustic pressure. Under these conditions, the heat release rate response of the flame to the given velocity fluctuations can be expressed using the $n-\tau$ model as follows \cite{Polifke2001,murugesan2018onset}.

\begin{equation}
    \frac{\hat{\dot{Q}}(\omega)}{\bar{\dot{Q}}} = n e^{-i\omega \tau_{conv}} \frac{\hat{v}_1(\omega,y_f)}{\bar{v}_1}
    \label{Eq_n_tau_model}
\end{equation}
Here, $n$ and $\tau_{conv}$ denotes the interaction index and convective time delay. The convective time delay is calculated as discussed below from the mean CH$^*$ chemiluminescence image and mean axial flow velocity at the dump plane.

Combining the acoustic pressure and velocity across the elements (swirl cup, combustion chamber and nozzle) along with the jump conditions and boundary conditions the dispersion relation is obtained and expressed as follows.

\begin{multline}
    (1 + e^{i \omega \tau_1}) \left( 
    \bar{a}_2 \left[
        e^{-i \omega \tau_3} (-1 + i k_3 \sqrt{S_3}) 
        + e^{-i \omega \tau_2} (1 + i k_3 \sqrt{S_3})
    \right] \right. \\
    \left.
    [1 + e^{-i \omega \tau_2}] + i \omega \sqrt{S_3} (e^{-i \omega \tau_2} - e^{-i \omega \tau_3}) 
    [1 - e^{-i \omega \tau_2}] 
    \right) \\
    + \frac{\bar{\rho}_2 \bar{a}_2 S_1}{\bar{\rho}_1 \bar{a}_1 S_2} 
    (e^{i \omega \tau_1} - 1) 
    \left( 1 + \frac{(\gamma - 1)}{\gamma \bar{p} S_1} 
    \frac{\bar{\dot{Q}}}{\bar{v}_1} n e^{-i \omega \tau_{conv}} \right) \\
 \left( 
    \bar{a}_2 \left[
        e^{-i \omega \tau_3} (-1 + i k_3 \sqrt{S_3}) 
        + e^{-i \omega \tau_2} (1 + i k_3 \sqrt{S_3})
    \right] 
    [1 - e^{-i \omega \tau_2}] \right. \\
    \left. + i \omega \sqrt{S_3} 
    (e^{-i \omega \tau_2} - e^{-i \omega \tau_3}) 
    [1 + e^{-i \omega \tau_2}]
    \right) = 0
    \label{eq_disp_rel}
\end{multline}

\begin{figure}[t]
\centering
\includegraphics[width=0.48\textwidth]{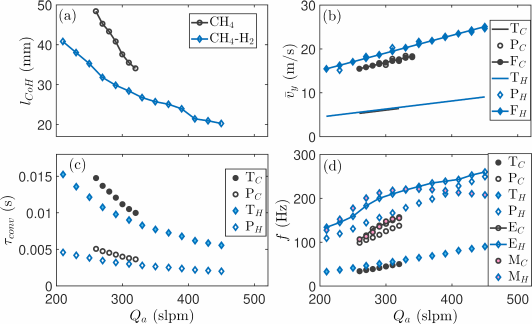}
\caption{(a) Center of heat release ($l_{CoH}$) of mean CH$^*$ chemiluminescence used in estimating the flame convective time delay \cite{kim2010effect}. (b) Mean axial velocity ($\bar{v}_y$), (c) convective time delay ($\tau_{conv}$), and (d) instability frequency ($f$), respectively. In panels (b-d), the notations T: theory, P: PIV experiments, F: linear least square fit, E: experiment (from Fig. \ref{fig_Prms_Fdom_SPL}b), and M: low order model (from Eq. \ref{eq_disp_rel}), respectively. The black and blue colors ---also the subscripts $[~]_C$ and $[~]_H$ in figure legends--- correspond to the CH$_4$ and CH$_4$-H$_2$ mixtures, respectively.} \label{fig_COM_Tau}
\vspace{-4mm}
\end{figure}

The time delays are $\tau_1=2l_s/\bar{a}_1, ~\tau_2=2 l_c/\bar{a}_2$ and $\tau_3=2(l_c+l_n)/\bar{a}_3$. The flow and geometric properties are given in Table \ref{Tab1_flow_geom}. Equation (\ref{eq_disp_rel}) is solved using MATLAB. The real and imaginary parts of $\omega$ denote the angular frequency and growth rate. This study particularly focuses on the real part of $\omega$ and shown in Fig. \ref{fig_COM_Tau}(d) with filled symbols. In addition to Eq. (\ref{eq_disp_rel}), the observed dominant instability frequency can be calculated using reactant mixture convective time scale.

The calculation of convective time delay ($\tau_{conv}$) in Eq. (\ref{Eq_n_tau_model}) is carried out as discussed below. From the fuel injection location to the tip of the flare ($l_s$) is fixed and 30 mm. Further, from flare tip to the center of heat release (CoH) location varies depending on the fuel composition and the airflow rate. The location of CoH ($l_{CoH}$) is calculated using CH$^*$ mean flame images \cite{kim2010effect} and the values are shown in Fig. \ref{fig_COM_Tau}(a). Figure \ref{fig_COM_Tau}(a) shows a reduction in CoH location of the flame as the air-flow rate increases. This could be attributed to enhanced mixing of the fuel and air caused by the increased flow turbulence and swirl intensity. The difference between methane and hydrogen-methane mixture is attributed to an increased total mixture velocity due to more fuel flow rate and higher reactivity and diffusivity of hydrogen. Note that, however, we observed nearly monotonic reduction in the heat release rate center with flow rate for both fuels. This also shows that for higher airflow rates the flame becomes more compact and close to the flare tip for this injector until flame blowout is reached. Furthermore, the methane fuel showed a significant reduction in $l_{CoH}$ with airflow rate compared to the methane-hydrogen mixture attributed to the low extinction strain rate of the methane-air mixture.

The mean axial velocity of the incoming mixture is estimated from the total mass flow through the flare exit and also from the experimental measurements of axial (PIV) flow field. The variation in velocity with airflow rate is shown in Fig. \ref{fig_COM_Tau}(b) along with their least square linear fit. Since the vortex breakdown occurs near the flare exit (refer to Fig. \ref{fig_PIV_Mean_vel} top row) and the average velocity estimated based on the mass flow rate near the exit plane (refer to the solid lines in Fig. \ref{fig_COM_Tau}b) is different from the fresh mixture convection velocity from the experiments (roughly 3 times lower). Hence, the $\bar{v}_y$ is chosen at the jet core near the exit plane. Note that the PIV experiments are performed for the selected cases and further fitted with the least square method (refer to the labels F$_c$ and F$_H$ in Fig. \ref{fig_COM_Tau}b for least square fit). Nearly a linear increase in flow velocity in Fig. \ref{fig_COM_Tau}(b) for both CH$_4$ and CH$_4$-H$_2$ air mixtures are observed. This is due to an increase in airflow rate while keeping the fuel flow rate constant. The difference between both cases is attributed to an increased fuel flow rate of the hydrogen-methane mixture to maintain the same thermal power (16 kW) and relative position of the IRZ with respect to flare exit.

The mixture convective time delay is estimated using the following expression.
\begin{equation}
 \tau_{conv}=(l_{s}+l_{CoH})/\bar{v}_y   
 \label{eq_tau_conv}
\end{equation}
The calculated $\tau_{conv}$ values are shown in Fig. \ref{fig_COM_Tau}(c). The $\tau_{conv}$ is estimated using both average flow velocity from theory and PIV experiments as discussed above. The $\tau_{conv}$ from theoretically estimated velocity is much larger compared with the estimation using PIV experiments. For instance, refer to the filled and open black circles in Fig. \ref{fig_COM_Tau}(c). The dominant instability frequency estimated from the acoustic pressure shown in Fig. \ref{fig_Prms_Fdom_SPL}(b) are reproduced using the $\tau_{conv}$ as discussed below. 

As mentioned in supplementary section \ref{sec_prim_charac}, it is to be noted that the $f_{dom}$ in Fig. \ref{fig_Prms_Fdom_SPL}(b) scales approximately linear with airflow rate. This feature is attributed to the ITA modes in combustion instability \cite{hoeijmakers2014intrinsic,emmert2015intrinsic}. Based on the $\pi-$criterion, the ITA mode frequency ($f_{ITA}$) is expressed as follows \cite{hoeijmakers2014intrinsic,emmert2015intrinsic,silva2023intrinsic}.
\begin{equation}
    f_{ITA}=\frac{1}{2 \tau_{conv}}  = \frac{\bar{v}_y}{2(l_{s}+l_{CoH})}
    \label{eq_ITA_freq}
\end{equation}
The ITA mode frequency estimated using Eq. (\ref{eq_ITA_freq}) is shown in Fig. \ref{fig_COM_Tau}(d) for both CH$_4$ and CH$_4$-H$_2$ mixtures, for instance, refer to T$_c$, P$_c$, T$_H$, and P$_H$. The symbols with connected lines correspond to the dominant instability frequency from experiments. We found a nearly good match between the estimation from Eq. (\ref{eq_ITA_freq}) and experiments at low amplitudes. For flow rates corresponding to the relatively large acoustic pressure amplitude the experimental values slightly deviate from linear variation corresponding to Eq. (\ref{eq_ITA_freq}). This is attributed to the nonlinear dependence of ITA mode instability frequency with amplitudes \cite{mohan2020nonlinear}. Furthermore, we numerically calculate the dominant mode frequency using the dispersion relation Eq. (\ref{eq_disp_rel}) and shown using symbols filled with red colors in Fig. \ref{fig_COM_Tau}(d). The LOM accurately predicts  ITA mode frequency for methane for all airflow rates considered. However, this is not the case for the methane-hydrogen mixtures. This could be improved by incorporating a more accurate description of the flame response \cite{mohan2020nonlinear,wildemans2023nonlinear}. From the above discussions, it is clear that the high-shear radial injector devised here for the examined operating conditions exhibits ITA mode combustion instability. In the following subsections, we provide insight into the associated flame and flow dynamics using DMD modes.

\subsection{Flame dynamics \label{sec_Flm_DMD}} 
\begin{figure}[h!]
    \centering
    \hspace*{-3mm}
    \includegraphics[width=0.48\textwidth]{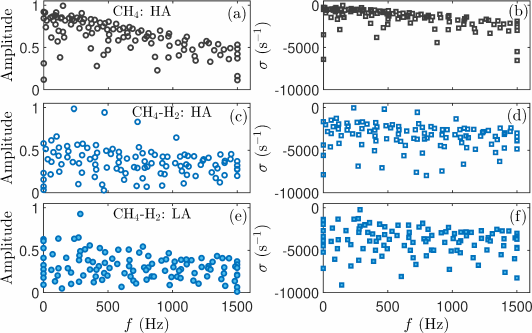}
    \caption{ DMD mode (right column) amplitude and (left column) growth rate ($\sigma$) associated with OH$^*$ chemiluminescence high-speed flame images for (a,b) CH$_4$: HA, (c,d) CH$_4$-H$_2$: HA, and (e,f) CH$_4$-H$_2$: LA, respectively. The open and filled symbols denote HA and LA cases, respectively. The circle and square markers denote DMD amplitude and growth rate.}
    \label{fig_DMD_amp_growth_flame}
\end{figure}

The combustion dynamics of the ITA mode from high-speed images are scrutinized using DMD modes to illustrate the heat release rate response. Figure \ref{fig_DMD_amp_growth_flame}(a) illustrates the DMD mode amplitude for the HA case of methane combustion. The dominant fundamental mode frequency is observed at approximately 149 Hz, which carries the most substantial contribution to the overall flame dynamics. In addition to this fundamental mode, first harmonics is present at around 293 Hz. The presence of harmonic structures suggests that the flame behaves nonlinearly. On the other hand, the higher amplitude case of the methane-hydrogen mixture shown in Fig. \ref{fig_DMD_amp_growth_flame}(c) displays the DMD mode amplitude comprising multiple higher harmonics along with the fundamental mode. The presence of higher harmonics suggests stronger nonlinear interactions and enhanced flame response due to hydrogen addition \cite{indlekofer2021effect}. These nonlinear interactions diminishes in the LA case of methane-hydrogen mixture as shown in Fig. \ref{fig_DMD_amp_growth_flame}(e), only fundamental mode persists. No dominant mode frequency is observed in heat release rate of methane LA case. This suggests that the combustion dynamics do not exhibit significant coherent flame surface oscillations in this condition. As a result, we do not report the DMD mode spectrum associated with the fundamental ITA frequency for this case. From the above, it is clear that the flame surface fluctuations induce higher harmonics in DMD spectrum for large amplitudes and their strength diminishes as we approach the low-amplitude acoustic oscillations. It is interesting to note that the harmonics are relatively more prominent in heat release rate spectrum than the acoustic pressure spectrum, shown in supplementary \ref{sec_prim_charac}.

The growth rates of DMD modes discussed above are shown in Fig. \ref{fig_DMD_amp_growth_flame} second column, respectively. The growth rate ($\sigma$) associated with the fundamental mode exhibits zero. This implies that the heat release fluctuations associated with these modes feature limit cycle oscillations. Further, this shows that the oscillations have reached a steady-state condition where energy responsible for instability growth and decay balance each other, typical of the self-sustained oscillator. 

\begin{figure}[h!]
    \centering
    \hspace*{-5mm}
    \includegraphics[width=0.49\textwidth]{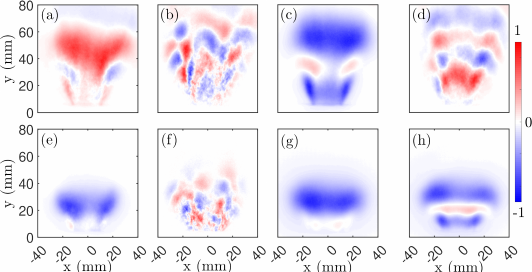}
    \caption{ OH$^*$ chemiluminescence high-speed flame imaging DMD modes associated with the fundamental thermoacoustic mode ($f_{dom}$) for (a,b) CH$_4$: LA, (c,d) CH$_4$: HA, (e,f) CH$_4$-H$_2$: LA, and (g,h) CH$_4$-H$_2$: HA, respectively. Panels (a,c,e,g) and (b,d,f,h) correspond to the fundamental and first harmonic of thermoacoustic modes, respectively.}
    \label{fig_DMD_mode_flame}
    \vspace{-3mm}
\end{figure}

Figure \ref{fig_DMD_mode_flame} presents the first two DMD modes for pure methane (top row) and methane-hydrogen mixture (bottom row) for both LA and HA cases. These DMD modes provide insight into the spatial distribution of heat release rate fluctuations among different modes and frequencies. As discussed previously, in the LA case of pure methane combustion, no dominant mode was observed due to the absence of significant coherent oscillatory structures compared to background level. This is visually evident in Fig. \ref{fig_DMD_mode_flame}(a) and \ref{fig_DMD_mode_flame}(b), where the first and second DMD modes do not exhibit distinct spatial features. The lack of clear modal structures illustrates the flame is in a more stable regime, with weak or negligible thermoacoustic interactions. Figures \ref{fig_DMD_mode_flame}(c) and \ref{fig_DMD_mode_flame}(d) illustrate the first and second DMD modes for the HA case of pure methane combustion. In contrast to the LA case, distinct spatial patterns are clearly visible. In the first mode, the flame exhibits a positive intensity region near the stagnation point located just after the recirculation zone separated by the negative heat release rate fluctuations. This indicates possible flow field influence on heat release, which likely contribute to the thermoacoustic feedback mechanism as discussed in the following subsection. Further, the second mode do not show a clear mode structure and appears to be distributed due to its relatively low contribution.


For the LA case of the methane-hydrogen mixture, Fig. \ref{fig_DMD_mode_flame}(e) and \ref{fig_DMD_mode_flame}(f) exhibit similar characteristics to the pure methane case. The absence of strong spatially coherent structures in both DMD modes suggests that the flame oscillations remain weak, and no dominant thermoacoustic mode is distinctly observed. This further confirms that at lower amplitudes, both fuel compositions exhibit relatively low amplitude heat release rate oscillations in the order of background noise level leading to nominal thermoacoustic feedback. In the HA case of the methane-hydrogen mixture, the dominant DMD modes exhibit a distinctly strong spatial distribution compared to all other cases analyzed. The heat release rate fluctuations are notably altered by cyclic positive and negative values. Unlike HA methane case where positive heat release rate distribution separate negative distribution on either side, methane-hydrogen case manifests cyclic downstream convection. In the second mode shown in Fig. \ref{fig_DMD_mode_flame}(h), because of the higher harmonic nature both positive and negative values can be seen compared to Fig. \ref{fig_DMD_mode_flame}(g).

\begin{figure}[h!]
    \centering
    \includegraphics[width=0.49\textwidth]{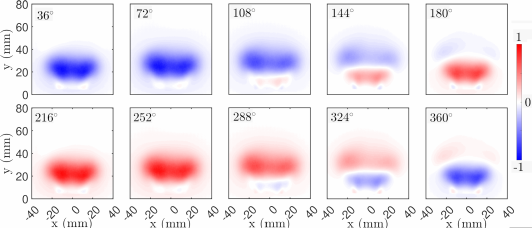}
    \caption{DMD mode of CH$_4$-H$_2$:HA (Fig. \ref{fig_DMD_mode_flame}g) over different phase angles illustrating the ITA mode heat release rate evolution.} 
    \label{fig_DMD_flame_H2CH4_US}
\end{figure}

Figure \ref{fig_DMD_flame_H2CH4_US} presents the phase evolution of OH* chemiluminescence distribution for the hydrogen-methane fuel mixture. The phase evolution of the dominant mode is illustrated using ten equally spaced angles. This allows for clear visualization of the temporal and spatial evolution of the maximum and minimum heat release rate regions of the OH$^*$ chemiluminescence spatial distribution. At 36$^\circ$, the negative OH$^*$ intensity is observed throughout with a maximum value near zero occurring close to the flare exit. However, as the phase evolves further, the maximum in OH$^*$ intensity starts appearing uniformly across the IRZ near the flare exit and further traversing the IRZ over the instability cycle. For instance, the near-zero values in phase angles $360^\circ$ to $72^\circ$, begins to detach from the flare exit at 108$^\circ$ and convected downstream, simultaneously interacting with the incoming fresh reactants. At the same phase angle, the weak minimum emerges at the shear layer. By 216$^\circ$, the maximum in HRR region expands significantly, covering almost the entire flame structure with a localized minimum in HRR region persisting near the outer shear layer. At this instant, the heat release rate fluctuations reach their maximum intensity. Through similar process by 360$^\circ$, the minimum in HRR region grows and completely dominates the downstream region of the flame, while the maximum in HRR convects downstream  and dissipates. Unlike the methane case, where the maximum in HRR region predominantly forms downstream of the recirculation zone (in the wake) and propagates into the IRZ through inner shear layer as discussed in the supplementary section \ref{CH4_HA_flame_DMD}, the hydrogen-methane mixture exhibits an initial formation of maximum in HRR in the root of the inner shear layer near the flare and traverses nearly half of the IRZ before reaching their extremum OH$^*$ intensity values. This difference in heat release rate fluctuations driving mechanism could be attributed to differences in average strain rate between these two mixtures shown in Fig. \ref{fig_PIV_Mean_vel}(f,h) and their relative unburnt mixture diffusivity. In the following subsection, we scrutinize the axial flow field using DMD modes to elucidate the underlying driving mechanisms.

\subsection{Flow dynamics \label{sec_PIV_DMD}} 

\begin{figure}[h!]
    \centering
    \hspace{-5mm}
    \includegraphics[width=0.5\textwidth]{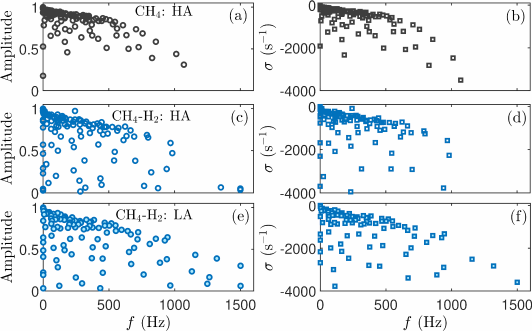}
    \caption{ DMD mode (first column) amplitude and (second column) growth rate associated with axial velocity ($v_y$) for (a,b) CH$_4$: HA, (c,d) CH$_4$-H$_2$: HA, and (e,f) CH$_4$-H$_2$: LA, respectively. The open and filled symbols denote HA and LA cases, respectively. The circle and square markers denote DMD mode amplitude and growth rate.}
    \label{fig_DMD_amp_growth_PIV}
\end{figure}

Figure \ref{fig_DMD_amp_growth_PIV} illustrates the DMD mode amplitudes (first column) and the corresponding growth rates (second column) for the axial velocity field similar to Fig. \ref{fig_DMD_amp_growth_flame}. In contrast to the DMD analysis of OH$^*$ chemiluminescence images, which reveals distinct dominant thermoacoustic modes, the velocity field does not exhibit significant modal amplitudes in most cases at the dominant ITA mode observed in acoustic pressure. However, in the HA hydrogen-methane mixture case, a pronounced peak emerges at a fundamental frequency as can be seen in Fig. \ref{fig_DMD_amp_growth_PIV}(c,d). This indicates a strong coupling between the flow field oscillations and the OH$^*$ chemiluminescence intensity variations, suggesting a direct interaction between the coherent structures and the heat release rate fluctuations. In the remaining cases, including HA methane Fig. \ref{fig_DMD_amp_growth_PIV}(a,b) and the LA hydrogen-methane mixture Fig. \ref{fig_DMD_amp_growth_PIV}(e,f), no dominant peak is observed in the velocity field, implying the absence of significant hydrodynamic-acoustic coupling. Based on the triple decomposition of the instantaneous velocity field, we found that the coherent component of the velocity field is much less than the turbulent intensity, at least $\mathcal{O}(10^{-2})$ less (shown in Table \ref{Tab_vor_shedd} of section \ref{vor_dyn}). In addition, the growth rate shown in the second column further corroborates, as only the methane-hydrogen HA case features a near-zero growth rate. The other cases feature a negative growth rate.

\begin{figure}[h!]
    \centering
    \hspace*{-5 mm}
    \includegraphics[trim={0 7cm 0 15mm},width=0.49\textwidth]{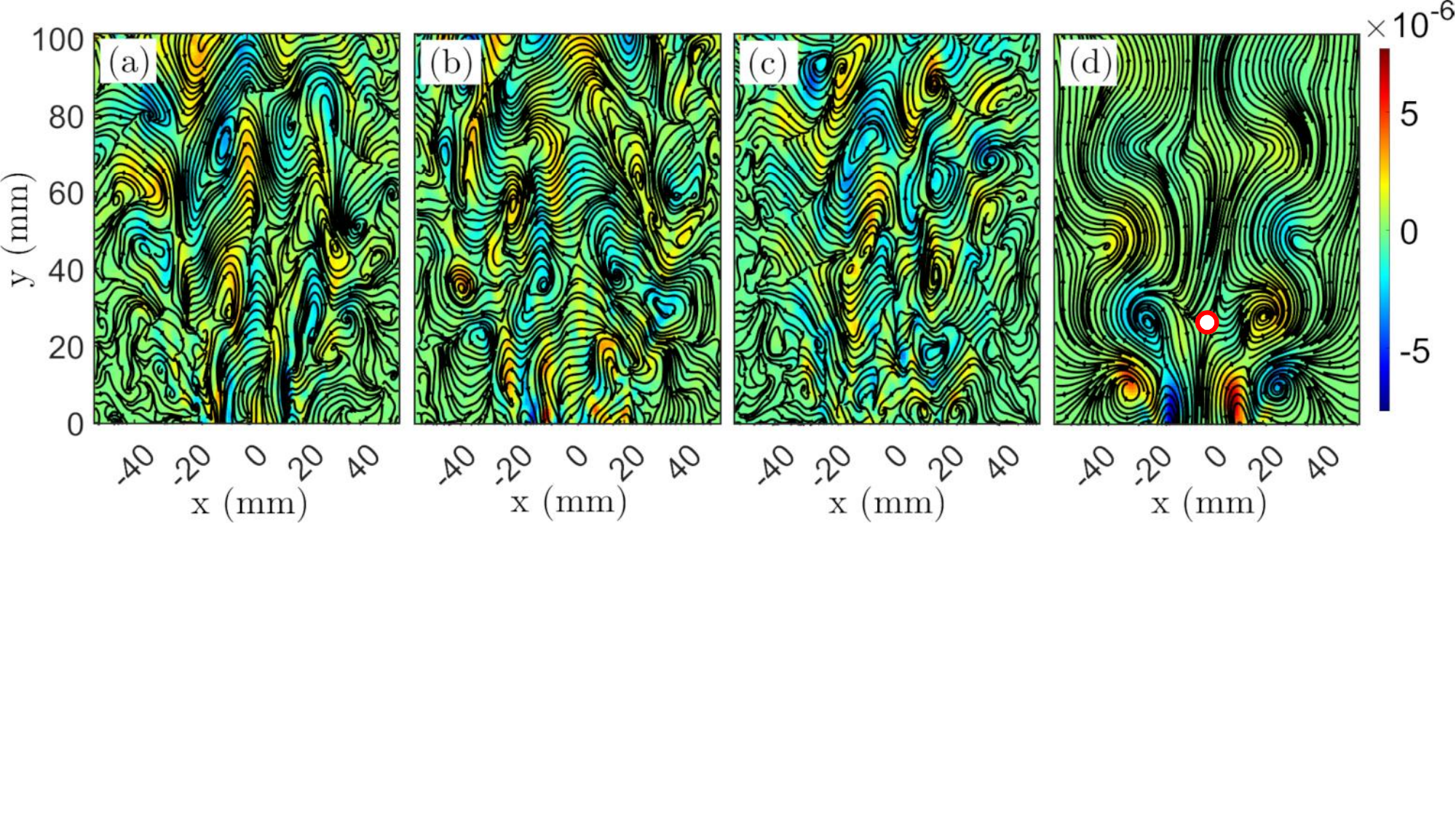}
    \caption{Vorticity field (in 1/s) DMD modes associated with the fundamental ITA mode ($f_{dom}$) for (a) CH$_4$:LA, (b) CH$_4$:HA, (c) CH$_4$-H$_2$:LA, and (d) CH$_4$-H$_2$:HA, respectively. The white dot enclosed with red colour circle indicates the stagnation point in panel (d).}
    \label{fig_DMD_mode_PIV}
\end{figure}

\begin{figure}[h!]
    \centering
    \hspace{-5 mm}
    \includegraphics[trim={0 10mm 0 0},width=0.49\textwidth]{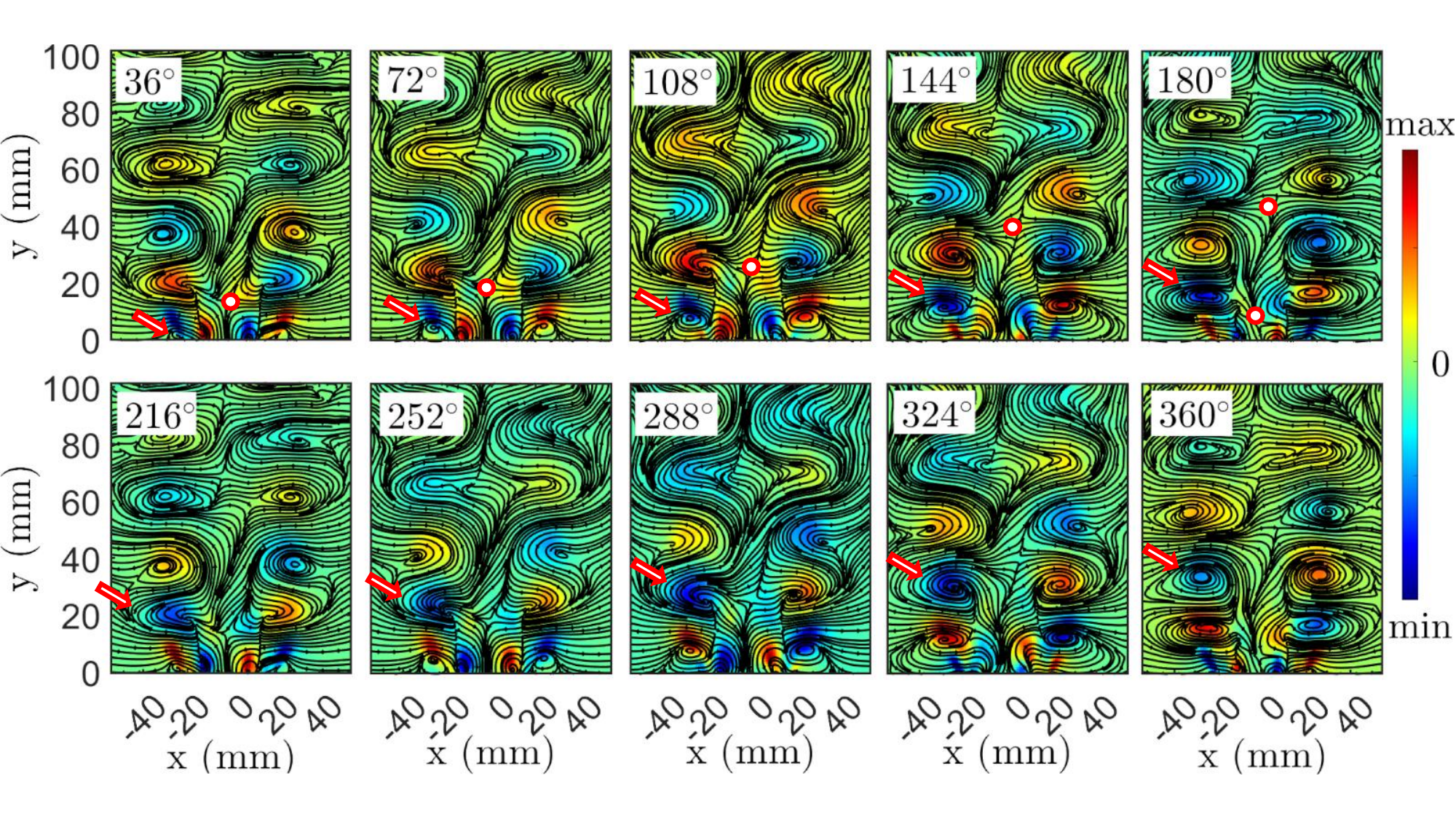}
    \caption{Vorticity field (in 1/s) DMD mode over different phase angles associated with CH$_4$-H$_2$:HA shown in Fig. \ref{fig_DMD_mode_PIV}(d). The white dot enclosed with red colour circle indicates the stagnation point and red arrow denotes the periodic vortex shedding.}
    \label{fig_DMD_PIV_H2CH4_US}
    \vspace{-4mm}
\end{figure}

Figure \ref{fig_DMD_mode_PIV} presents the vorticity field corresponding to the dominant thermoacoustic mode coherent structures extracted from the DMD analysis for both pure methane and hydrogen-methane mixture fuel cases. In the HA methane-hydrogen case, a well-defined coherent structure is observed, as depicted in Fig. \ref{fig_DMD_mode_PIV}(d). 
The observed vortex pairs exhibit a symmetric spatial distribution with respect to the vertical axis, indicative of a dominant instability mode governing the flow dynamics. These structures originate primarily from the outer recirculation zone and propagate downstream, as shown in Figure \ref{fig_DMD_mode_PIV}(d), reinforcing the presence of large-scale coherent vortices that modulate the thermoacoustic oscillations. The strong coupling between the hydrodynamic and thermoacoustic instabilities in this case suggests that the coherent structures play a crucial role in amplifying the limit-cycle oscillations. The periodic vortex shedding observed in this regime further confirms the presence of hydrodynamic instability mechanisms, which drive the self-sustaining oscillatory behaviour \cite{OCONNOR2023}.

In contrast, the HA case of pure methane combustion does not exhibit a well-defined recirculation-induced coherent vortex structure as shown in Fig. \ref{fig_DMD_mode_PIV}(b). Instead, weaker vortical structures are present which primarily convected downstream along both the inner and outer shear layers. This suggests that while hydrodynamic perturbations exist, their interaction with the flame dynamics is not as strongly coupled as in the hydrogen-methane HA case. For the lower amplitude cases of both pure methane and hydrogen-methane mixtures as shown in Fig. \ref{fig_DMD_mode_PIV}(a,c), no dominant coherent structures are detected with the fundamental thermoacoustic mode. The absence of significant coherent vorticity structures in these cases implies that the flow remains relatively stable, with minimal or no vortex shedding. This further confirms that at lower oscillation amplitudes, the hydrodynamic perturbations do not sufficiently interact with the flame to drive strong thermoacoustic coupling and vice versa, resulting in negligible hydrodynamic instability effects \cite{saurabh2014swirl,OCONNOR2023}.

To further analyze the dynamic behavior of the fundamental thermoacoustic mode flow field, the phase evolution of DMD-extracted coherent structures is examined. 
Figure \ref{fig_DMD_PIV_H2CH4_US} presents the phase evolution of coherent structure for the hydrogen-methane mixture at the fundamental ITA mode. The location of the stagnation point at each phase angle is marked with a white dot circled with red colour until the 180$^\circ$ for clarity (top panel). At 36$^\circ$, the coherent structures originate predominantly from the outer shear layer, as indicated by the white arrow with red outline. By 108$^\circ$, these vortices have fully developed and exhibit strong rotational motion. At 144$^\circ$, the vortices reach their peak strength, leading to a momentary suppression of the recirculation bubble. This suppression mechanism is further amplified at 180$^\circ$, where the existing stagnation point is displaced, leading to the formation of a new stagnation point further upstream. Between 216$^\circ$ and 360$^\circ$, the vortices convected towards the downstream region, repeating the stagnation point movement observed in the first half-cycle. This periodic cycle demonstrates that, despite the secondary swirler contributing a lower mass flow rate compared to the primary swirler, the outer shear layer vortices initiate the coherent vortex formation and subsequently fed by the inner swirl jet leading to peak vorticity value followed by detachment and convection downstream. Therefore, the flow dynamics is primarily governed by the initial generation of coherent structure from the outer shear layer whereas in pure methane case, the inner recirculation vortices exhibit a rapid downstream convection compared to the outer recirculation vortices as discussed in the supplementary section \ref{CH4_HA_flow_DMD}. This observation highlights the non-reacting flow from the secondary swirlers critical role in dictating the large-scale coherent structures within the combustor, thereby influencing the overall heat release rate dynamics and subsequent dynamic coupling. Insight into the vorticity dynamics relating to the circulation and its convective motion along the jet axis further discussed in the following subsection using a simplified vorticity dynamics model. 

\subsection{Vorticity dynamics with simplified model \label{vor_dyn}}
\begin{figure}[h!]
    \centering
    \includegraphics[width=0.48\textwidth]{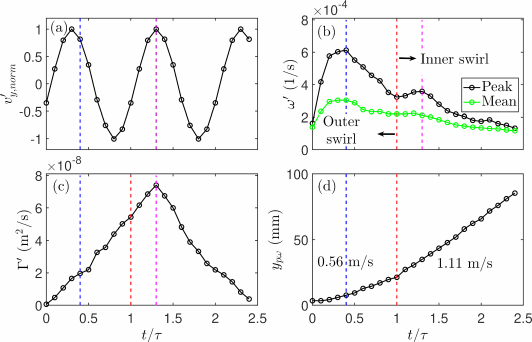}
    \caption{(a) Normalized axial velocity fluctuations, (b) peak vorticity, (c) circulation, and (d) location of peak vorticity for  CH$_4$-H$_2$: HA case associated with Fig. \ref{fig_DMD_PIV_H2CH4_US}. The vertical dashed lines in panel (b) denote local maxima (blue, magenta) and minimum (red) in peak vortices for clarity.}
    \label{fig_Circulation}
\end{figure}

Based on the above discussions, we found a strong correlation at the dominant ITA mode frequency between the flow field, OH* chemiluminescence oscillations, and acoustic pressure for the methane–hydrogen fuel at HA case. 
To investigate this coupling further, the flow field shown in Fig. \ref{fig_DMD_PIV_H2CH4_US} was examined in detail, particularly focusing on the vortex formation and impact of outer and inner swirl flows on vorticity dynamics. The results are presented in Fig. \ref{fig_Circulation}, where all data are plotted against the non-dimensional time scale $t/\tau$, with $t$ the time and $\tau$ the period corresponding to the fundamental ITA frequency.
Figure \ref{fig_Circulation}(a) shows the normalized fluctuating component of the axial velocity extracted from the region near the flare exit. A pronounced peak occurs at $t/\tau=0.3$, indicating a surge in upward momentum associated with the initial vortex formation and detachment from the outer swirl flow. This fluctuation decays and reaches a local minimum at $t/\tau=0.8$, before increasing again, suggesting a periodic nature of the vortex formation and shedding processes. 

The average vorticity and circulation are identified in this study by integrating isolated vorticity fields having a threshold value above 10\% of peak vorticity. In the present study, the choice of vorticity ($\omega^\prime$) threshold values affects the vortex formation, shedding from the outer swirl jet and subsequent interaction with the inner swirl jet before downstream convection quantitatively, for instance, refer to \cite{krueger2006formation,wang2021dynamics}. However, the underlying vortex dynamics mechanism is preserved for the threshold value considered here. In Fig. \ref{fig_Circulation}(b-d), the blue, red and magenta coloured vertical lines represent the vortex shedding from the outer swirl jet, demarcation of outer and inner swirl jets influence on vortex convection and vortex detachment from the inner swirl jet, respectively. Further Fig. \ref{fig_Circulation}(b) depicts peak and average vorticity values by the black and green colours, respectively. Figures \ref{fig_Circulation}(c,d) show circulation and location of peak vorticity along the axial direction. 

In Fig. \ref{fig_Circulation}(b), a peak value at $t/\tau=0.4$ occurs, which is close to the axial velocity peak observed in Fig. \ref{fig_Circulation}(a) illustrating the vortex shedding event. 
After shedding from outer swirl jet, the peak and average vorticity values decrease for a while due to dissipation until the inner swirl jet feeds the vorticity. During this feeding process, the vortex spreads out spatially and its peak value decreases significantly. This dissipation of vorticity is clearly discernible in the peak vorticity value compared to its mean. However, the circulation shown in Fig. \ref{fig_Circulation}(c) grows linearly having nearly the same slope until complete detachment from the inner swirl jet occurring at $t/\tau=1.30$. This $t/\tau=1.30$ coincides with the peak axial velocity fluctuations associated with the next cycle. Therefore, a momentary increase in axial velocity fluctuations leads to vortex shedding from both outer and inner shear layers. Further, a relatively mild increase in circulation rate is observed from inner swirl jet compared to the outer swirl jet. During the slope change in Fig. \ref{fig_Circulation}(c) occurring between $t/\tau=0.4-0.7$, circulation level deviates slightly from linear increase attributed to the dissipation. As shown in Fig. \ref{fig_Circulation}(d), the convection of vorticity occurs in two phases influenced by the outer and inner swirl jets. This convection velocity appears to be independent of the vortex shedding processes rather influenced by the feeding source and their relative contribution. Finally, in Fig. \ref{fig_Circulation}(c), the circulation decreases after detachment from the inner swirl jet attributed to the vortex dissipation by entrainment. It is interesting to point out that the non-reacting jet emanating from the outer swirl jet initiates the onset of vortex shedding. By tweaking the secondary jet passively the initial vortex formation can be suppressed or the total vortex generation can be nullified leading to a relatively stable operation hydrogen combustion.

The relative magnitudes of turbulent and coherent fluctuations in the flow field determines the formation and evolution of vortices, for instance, refer to \cite{karmarkar2023interaction}. Therefore, we estimate the coherent and turbulent axial velocity contributions from the experiments and summarize it in Table \ref{Tab_vor_shedd} for clarity. Apart from the methane-hydrogen HA case, the remaining cases feature two orders of low coherent fluctuations compared to their turbulent counterpart. Therefore, no clear vortex shedding is observed in the experiments. To further clarify the above vortex shedding events across ITA mode frequency observed in the experiments, we use the vorticity dynamics model and relate it to all the cases examined.

\begin{table}[t]
\centering
\hspace*{-4mm}
\resizebox{75mm}{!} 
{ 
\begin{tabular}{|c| c| c| c| c| c| c|}
\hline
   Case & $Q_a$ (slpm) & Mixture & $|\hat{q}_c|/\bar{q}~ \%$ &$|\hat{v}_{c}|/\bar{v}_y$ \% & $ {v}_{t,rms}/\bar{v}_y$  \% & Vortex Shedding \\ \hline
 1 & 300 & CH$_4$ & 18.15 & 0.96 & 17.84 & No\\ 
 2 & 320 & CH$_4$ & 3.13 & 0.13 & 19.57 &  No\\   
  3 & 390 & CH$_4$-H$_2$ & 64.50 & 11.21 & 18.40 &  Axisymmetric\\
  4 & 450 & CH$_4$-H$_2$ & 5.10 & 0.83 & 18.36 &   No
  \\ \hline
\end{tabular}
}
\caption{Relative contributions of coherent ($|\hat{v}_{c}|/\bar{v}_y$) and turbulent ($ {v}_{t,rms}/\bar{v}_y$) velocity fluctuations across different cases examined relating to the vortex dynamics.} \label{Tab_vor_shedd}
\vspace{-3mm}
\end{table}

\begin{figure}[h!]
    \centering
    \includegraphics[width=0.49\textwidth]{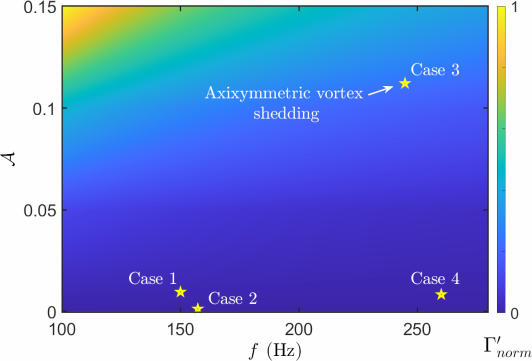}
    \caption{Normalized circulation from Eq. (\ref{eq_cir_fluc}) across dominant instability frequency and normalized coherent velocity amplitude with $\delta=0$. The star symbols denote the frequency and amplitude associated with the cases listed in Table \ref{Tab_vor_shedd}. }
    \label{fig_Circulation_1}
    \vspace{-4mm}
\end{figure}

The vortex formation mechanism function of velocity fluctuations can be explained following the previous studies \cite{wang2021dynamics}. The instantaneous circulation evolution can be expressed as follows.
\begin{equation}
    \frac{d \tilde{\Gamma} }{dt}= \frac{1}{2} \tilde{v}_y^2 \label{eq_mean_circ}
\end{equation}
Decomposing the instantaneous quantities into mean, coherent, and turbulence quantities ($\tilde{v}=\bar{v}+v_c^\prime+v_t^\prime$), the fluctuating circulation can be estimated from Eq. (\ref{eq_mean_circ}). The expression for circulation increments with time as follows assuming turbulence velocity fluctuations as uncorrelated white noise. This approximation of turbulent flow is not precise considering its nonlinear behaviour and associated diverse spatio-temporal scales. However, the contributions from the combustion noise assumed to take this particular form for simplification in this study.
\begin{equation}
    \Gamma^\prime = \frac{\bar{v}^2_y}{4} \left(\mathcal{A}^2 t_{circ} + \delta^2 \right) \label{eq_cir_fluc}
\end{equation}
In the above equation, $\mathcal{A}=|\hat{v}_{c}|/\bar{v}_y$, $t_{circ}$ and $\delta$ denote normalized coherent velocity fluctuations at the injector dump plane, circulation increment time $[0,T_{shed}]$, and normalized standard deviation (with $\bar{v}_y/\sqrt{2}$) of turbulent fluctuations, respectively. The normalized circulation from Eq. (\ref{eq_cir_fluc}) across dominant instability frequency and normalized velocity amplitude is plotted considering $\delta=0$ and shown in Fig. \ref{fig_Circulation_1}. The case 3 in Table \ref{Tab_vor_shedd} with a volumetric air flow rate of 390 slpm exhibits a significantly larger circulation level, exceeding 0.1, whereas all other cases remain below 0.01, irrespective of their oscillating frequency. This is inline with the DMD mode shapes depicted in Fig. \ref{fig_DMD_mode_PIV}. This clearly indicates that effective coupling among the three fields occurs only when the amplitude surpasses a critical threshold, which in this study is associated with strong axisymmetric vortex shedding of the flow field from large amplitude acoustic oscillations. 

\section{Conclusions \label{con}}
In this study, we scrutinized the ITA-driven combustion instability by elucidating the flow, flame, and acoustic coupling in detail. A model combustor comprising of a high shear injector having a 6:4 air split ratio between the inner and outer swirler is considered, and the combustor operates in a partially premixed mode. The pure methane and methane-hydrogen mixture having a fixed $50\%-50\%$ volume fraction are examined with a fixed thermal power of 16 kW and a Reynolds number range of 8000-19000. The acoustic pressure measurements from the combustion chamber show thermoacoustic fluctuations with a dominant instability frequency linearly increasing with airflow rate. Furthermore, the methane-hydrogen mixture exhibits wider operating conditions and relatively large amplitude acoustic fluctuations.

The average flame and flow features are scrutinized using high-speed OH$^*$ chemiluminescence images and high-speed PIV measurements. Pure methane exhibits relatively longer flames compared to the methane-hydrogen mixture. The low values of flow-induced strain rate in the inner swirl jet for the methane-hydrogen mixture along with its large value of extinction strain rate led to the compact and shorter flame. The flow field shows a relatively smaller IRZ despite having larger flow rates for the methane-hydrogen mixture. Because of the larger flame height and lower flow rate, pure methane shows a larger convective time delay compared to that of the methane-hydrogen mixture. A LOM is developed considering the swirl cup, combustion chamber, and exhaust nozzle section which incorporates the convective time delay. The LOM reproduced the experimentally observed dominant mode frequency substantiating the ITA nature of the observed combustion instability for hydrogen-methane fuel mixture. 

The DMD analysis of the high-speed flame images and flow fields provides possible coupling mechanisms pertaining to the ITA mode. In the pure methane case, the extremum in heat release rate fluctuations starts from the tip of the inner shear layer and spreads out to the wake and IRZ causing maximum and minimum values. On the other hand, in the methane-hydrogen cases, the extremum starts from the root of the inner shear layer and spreads out across the IRZ, causing large amplitude heat release rate fluctuations and dissipating significantly as it reaches the wake region. The large amplitude heat release rate fluctuations in methane-hydrogen cases are coupled with vortex dynamics in driving the ITA mode.

The onset of vortex shedding occurs from the outer shear layer fed by the outer swirl jet. The vortices convect and interact simultaneously with the inner swirl jet before shedding. The vortex shedding occurs from both outer and inner swirl jets at the instant of maximum velocity fluctuations. The spatial location of major heat release rate fluctuations coincided with the locations of initial phase of vortex formation, interaction with the inner swirl jet, and subsequent shedding. The relatively low amplitude cases do not show a clear vortex structure owing to the prominent background turbulence level. However, in this case the inner and outer vortex pairs do not interact with each other and convect at different velocities leading to minimal heat release rate fluctuations. A simplified vortex formation model is developed to clarify the vortex shedding behaviour by incorporating the coherent oscillation amplitude and ITA mode frequency from experiments. This model shows the large amplitude methane-hydrogen mixture circulation level is much larger than the other cases examined in this study, highlighting the influence of vortex dynamics on heat release rate. Further, this study shows the peak vorticity level arises from the outer swirl jet which can be passively modified to decouple the vortex shedding and hence neutralizing heat release rate fluctuations. This study provides valuable insight into the instability driving mechanism in multistage annular injectors used in modern gas turbines.

\section*{CrediT authorship contribution statement}

\textbf{SKT:} Conceptualization, Methodology, Investigation, Analysis, Writing – original draft, review \& editing. \textbf{BM:} Conceptualization, Methodology, Investigation,  Analysis, Writing – original draft, review \& editing. \textbf{SB:} Funding acquisition, Supervision, Writing – review \& editing.

\section*{Declaration of competing interest}

The authors declare none.

\section*{Acknowledgments}

The authors thank DRDO for financial support with grant number 1-C-5.  The authors acknowledge Dr. Pratikash Panda from the Aerospace Engineering Department of IISc Bengaluru for providing a high-speed IRO.

\section*{Supplementary material}   
\subsection{Dynamic Mode Decomposition \label{supl:DMD}}
The Dynamic Mode Decomposition (DMD) is applied to the OH$^*$ chemiluminescence flame images and PIV flow field to elucidate the physical mechanism associated with the instability modes \cite{schmid2010dynamic}. The DMD modes provide insight into the coherent structures specific to the instability frequency. For this purpose, the OH$^*$ chemiluminescence or flow field snapshots at $k^{th}$ time instant can be represented by $x_k$. The total number of snapshots considered is $N$ and $k=[1,2, \dots, N]$. The sequence of snapshots collected as column vectors separated by a sampling time ($dt=1/$ fps) is represented as $X_1=[x_1, x_2, \cdots x_{N-1}]$ and $X_2=[x_2, x_3, \cdots, x_N]$. These two sets of snapshot matrices are related through matrix $A$ as follows.

\begin{equation}
    X_2= A X_1
\end{equation}
The estimation of $A$ is performed using the reduced order approximation through the singular value decomposition \cite{kutz2016dynamic}. The singular value decomposition of $X_1$ after the reduced-order approximation, $X_1   \approx U \Sigma V^*$. Here, the superscript $[~]^*$ denotes the complex conjugate. Further, the $A$ is approximated using an operator $\tilde{A}= U^* X_2 V \Sigma^{-1}$. The eigenvalue and eigenvector of $\tilde{A}$ are related to $\tilde{A}W=\lambda W$. The real and imaginary parts of the eigenvalue ($\lambda$) of $\tilde{A}$ matrix are related to the growth rate and oscillating frequency, respectively. The $k-th$ complex frequency is expressed as $\Omega_k=log(\lambda_k)/dt$. Further, the DMD modes are estimated using the following expression.
\begin{equation}
    \Phi = X_2 V \Sigma^{-1} W.
\end{equation}
In addition, the reconstructed field can be expressed as follows.
\begin{equation}
    x(t) \approx \sum_{k=1}^{r}  \Phi_k e^{\Omega_k t} b_k
\end{equation}
Here, $b_k$ and $r$ denote the initial condition and truncation length, respectively. Note that the algorithm given in  \cite{kutz2016dynamic} is followed in estimating the DMD modes in this article. 

\subsection{Acoustic pressure characterization \label{sec_prim_charac}}

\begin{figure}[h!]
    \centering
    \hspace*{-3mm}
    \includegraphics[width=0.5\textwidth]{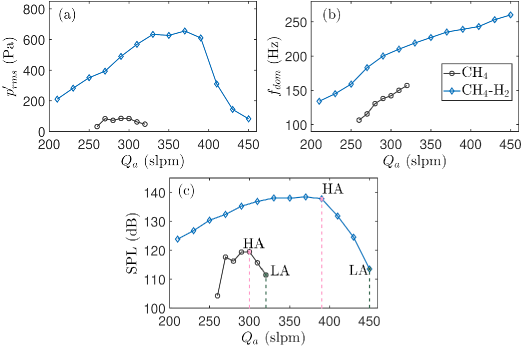}
    \caption{(a) Root Mean Square (RMS) value of acoustic pressure signal, (b) dominant instability frequency ($f_{dom}$), and associated (c) Sound Pressure Level (SPL) for various air flow rates ($Q_a$). The black and blue lines correspond to the methane ($Q_f=29.5$ slpm) and methane-hydrogen ($Q_f=43$ slpm) mixture, respectively. Pure methane exhibits relatively low amplitude and low-frequency oscillations compared with the methane-hydrogen mixture. The airflow rates marked with HA (high amplitude) and LA (low amplitude) highlighted using vertical lines in panel (c) are considered for further scrutiny.}
    \label{fig_Prms_Fdom_SPL}
\end{figure}

The differences in combustion dynamics between the methane and the methane-hydrogen mixture is examined in this section from the acquired acoustic pressure signals for various airflow rates. The airflow rates are varied such that both rich and lean blowout limits are reached for attached flame. Figure \ref{fig_Prms_Fdom_SPL}(a-c) represent the relationship between the airflow rate and the root mean square (RMS) value, dominant instability frequency, and associated sound pressure levels (SPL) of the acoustic pressure, respectively. The lines with black and blue colors correspond to the methane and methane-hydrogen mixture. For these experiments, the methane-hydrogen mixture was tested with an airflow rate increment of 20 slpm, while pure methane was tested with an increment of 10 slpm. The fuel flow rates were held constant at 29.5 slpm for methane and 43 slpm for the methane-hydrogen mixture, to achieve a constant thermal power of 16 kW in both the cases. Between these two fuel cases, the pure methane consistently exhibited lower amplitude RMS pressure signals and produced comparatively lower frequency oscillations. This difference underscores the stronger thermoacoustic response observed with the methane-hydrogen mixture which is influenced by the combustion properties of hydrogen, such as higher reactivity and flame speed along with combustor properties involving injector flow field, mode of operation, chamber acoustic field, heat release rate distribution and acoustic boundary conditions.

The RMS values of acoustic pressure fluctuations ($p^\prime_{rms}$) show distinct trends for both fuels. For all unstable cases of methane-hydrogen mixture, the RMS values tend to increase with airflow rate. However, as the system transitions to a stable operating regime, the $p^\prime_{rms}$ values decrease sharply. This behaviour is more pronounced for the methane-hydrogen mixture, where the transition between unstable and stable regimes results in significant drops in $p^\prime_{rms}$. This is because increasing the airflow rate increases the Reynolds number and enhances turbulent mixing, which alters the unsteady heat‐release fluctuations, approaching  premixed flame behavior \cite{kim2010response}. Furthermore, increasing airflow rate also alters the operating equivalence ratio which also results in modification in flame response \cite{balachandran2005experimental}. According to the Rayleigh criterion, when the heat‐release oscillations are in phase with pressure waves, acoustic energy is amplified, causing $p^\prime_{rms}$ to grow \cite{lieuwen2005combustion}. But when it moves towards the stable region, acoustic damping mechanisms dominate the driving, breaking the phase coherence required by the Rayleigh criterion. This leads to a strong decay in $p^\prime_{rms}$. 

However, the changes observed in the pure methane case are smaller and less pronounced, indicating a weaker thermoacoustic response compared to the methane-hydrogen mixture \cite{aesoy2020scaling}. But a discrete peak is observed in the Fourier spectrum. This is because, methane flames exhibit lower flame gain response and less phase sensitivity, so unsteady heat‑release fluctuations are smaller and the Rayleigh feedback loop is weak. This yields only modest broadband amplification and a low $p^\prime_{rms}$ rise with flow perturbations \cite{Masugi2025}. Furthermore, the RMS of acoustic pressure signal exhibits a similar qualitative variation with the airflow rate for both fuels in Fig. \ref{fig_Prms_Fdom_SPL}(a). The methane-hydrogen mixture exhibits wider operating conditions compared with pure methane. Similar observations of enhanced flammability limits for hydrogen addition was reported in the past \cite{schefer2003hydrogen,kim2009hydrogen}. In contrast, the larger thermoacoustic response observed in this combustor depends on burner configuration related to the Rayleigh phase and flame amplitude response, for instance, refer to Chterev and Boxx \cite{chterev2021effect} for time delay based Rayleigh phase criterion.

The dominant instability frequency ($f_{dom}$) in acoustic pressure spectrum are shown in Fig. \ref{fig_Prms_Fdom_SPL}(b). The instability frequency increases steadily with the airflow rate for both fuels. This trend appears nearly linear, suggesting a strong coupling between the air mass flow rate and the dominant instability frequency of the system. Further, it should be noted that nearly 50 Hz difference between the instability frequencies are observed between the two fuels examined. These differences could be attributed to the relative reduction in flame length, different burnt gas temperatures, and changes in the fuel flow rates. Further, the wide changes in instability frequency with flow rate also illustrates the observed instability mode is not related to duct acoustic modes, considering the low Mach number flows examined in this study \cite{ghirardo2018effect}.

The sound pressure levels (SPL) estimated at $|\hat{p}(f_{dom})|$ are shown in Fig. \ref{fig_Prms_Fdom_SPL}(c) which follows a trend similar to the RMS values in Fig. \ref{fig_Prms_Fdom_SPL}(a). The SPL increases with the air mass flow rate in the unstable regime but decreases sharply as the system approaches lean blow-off limit. Despite the lower amplitude and frequency of pure methane, its SPL still exhibits noticeable changes at dominant instability frequency, particularly at higher air mass flow rates. It is important to note the occurrence of rich blow-off under specific conditions influences the observations reported in Fig. \ref{fig_Prms_Fdom_SPL}(a-c). For pure methane, the rich blow-off was observed at air mass flow rates below 250 slpm, whereas for the methane-hydrogen mixture, this occurred at airflow rates below 200 slpm. This difference reflects the influence of hydrogen's lower flammability limits and higher flame speed, which lead to blow-off at relatively low airflow rate conditions when compared to pure methane \cite{dunn2011lean,beita2021thermoacoustic}. To understand the combustion dynamics further, the operating conditions marked with vertical dashed lines in Fig. \ref{fig_Prms_Fdom_SPL}(c) representative of High Amplitude (HA) and low Amplitude (LA) for both fuels are explored in detail.

\begin{figure}[h!]
    \centering
    \includegraphics[width=0.48\textwidth]{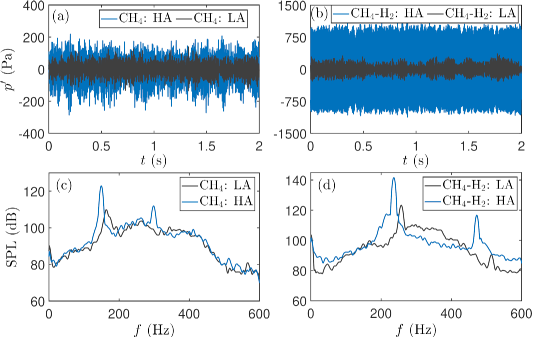}
    \caption{(a,b) Acoustic pressure time series and associated (c,d) SPL for (a,c) methane and (b,d) methane-hydrogen mixture. The blue and black lines illustrate the high amplitude (HA) and low amplitude (LA) acoustic pressure oscillations associated with the respective fuel-air mixtures highlighted in Fig. \ref{fig_Prms_Fdom_SPL}(c). The associated air flow rates ($Q_a$) for (a,c) 300 slpm (blue) and 320 slpm (black) and (b,d) 390 slpm (blue) and 450 slpm (black).}
    \label{fig_PrTS_SPL}
    \vspace{-5mm}
\end{figure}

Figure \ref{fig_PrTS_SPL} illustrates the acquired acoustic pressure signals for both methane and methane-hydrogen mixture. The black and blue lines in Fig. \ref{fig_PrTS_SPL} correspond to low amplitude (LA) and high amplitude (HA) operating conditions associated with the respective fuel flow rates highlighted in Fig. \ref{fig_Prms_Fdom_SPL}(c). Comparing the acoustic pressure signals in Figs. \ref{fig_PrTS_SPL}(a) and \ref{fig_PrTS_SPL}(b), the lower amplitude (LA) of the methane-hydrogen mixture closely resembles the higher amplitude (HA) of the pure methane pressure signals. The lower amplitude of methane pressure signals, however, is smaller, measuring 110 dB, and considered to be stable. However, this value is significantly more than the background noise level without combustion. In contrast, the higher amplitude signals of the methane-hydrogen mixture, reaching 142 dB, is significantly greater than all other cases shown in Fig. \ref{fig_PrTS_SPL}. This stark difference highlights the strong thermoacoustic coupling present in the methane-hydrogen mixture. The acoustic pressure signal in Fig. \ref{fig_PrTS_SPL}(b) for large amplitude exhibits limit cycle oscillation.

Figure \ref{fig_PrTS_SPL}(c) and \ref{fig_PrTS_SPL}(d) present the sound pressure levels in dB. Notably, the SPL for the lower amplitude (123.25 dB) of the methane-hydrogen mixture aligns closely with the higher amplitude SPL (122.8 dB) for pure methane at dominant frequency. This demonstrates that even at lower amplitudes, the methane-hydrogen mixture exhibits pronounced thermoacoustic behaviour. Further, it is interesting to point out that for large amplitudes the frequency is relatively lower and narrow compared to the low amplitude distributed peaks in Fig. \ref{fig_PrTS_SPL}(c,d). This shows the influence of amplitude dependency on dominant frequency is relatively less compared to the mean flow field \cite{mohan2020nonlinear}. As pointed out in the literature both suppression and amplification of thermoacoustic instability was observed with hydrogen addition to the natural gas \cite{kim2009hydrogen,chterev2021effect,beita2021thermoacoustic}. In summary, it is clear from Fig. \ref{fig_PrTS_SPL} that this test rig amplifies the thermoacoustic oscillations with hydrogen addition to methane. Further, the resulting instability frequency convectively scales with the airflow rate for both methane and methane-hydrogen mixtures, irrespective of their amplitude levels. In addition, methane-hydrogen mixture exhibits wider operating conditions with large amplitude thermoacoustic oscillations compared to pure methane. In the following subsection, the average flame and flow field will be discussed in detail. 

\subsection{Time averaged flame and flow features \label{avg_flame_PIV}}

\begin{figure}[h!]
    \centering
    \hspace*{-5mm}
    \includegraphics[width=0.51\textwidth]{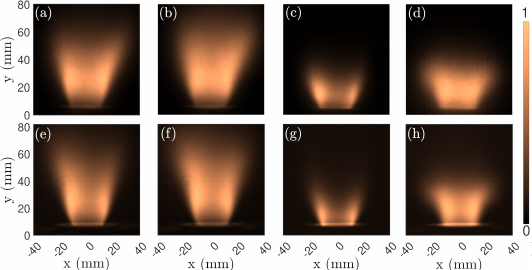}
    \caption{Average OH$^*$ (top) and CH$^*$ (bottom) chemiluminescence flame images corresponding to (a,e) CH$_4$: LA, (b,f) CH$_4$: HA, (c,g) CH$_4$-H$_2$: LA, and (d,h) CH$_4$-H$_2$: HA, respectively. The mean flame images are normalized with their respective maxima.}
    \label{fig_Mean_Flame_OH_CH}
\end{figure}

Figure \ref{fig_Mean_Flame_OH_CH} presents the mean OH$^*$ and CH$^*$ chemiluminescence images for both pure methane and methane-hydrogen mixtures under LA and HA conditions.  For pure methane combustion, the OH$^*$ chemiluminescence signal exhibits a lower flame height compared to the CH$^*$ chemiluminescence signal, as shown in Fig. \ref{fig_Mean_Flame_OH_CH} (a, b) and (e, f), respectively. This is because CH$^*$ remains trapped in the primary reaction zone and along the flame base, which extends higher. Its lower diffusivity and slower quenching allow the CH$^*$ signal to appear at greater heights, but OH$^*$ has higher diffusivity and shorter lifetime than CH$^*$, causing it to spread less far along the flame and attenuate faster with height \cite{YAN2023}. In addition, the relatively lower flame height of OH$^*$ indicates that the primary oxidation reactions in methane combustion occur over a compact region, with CH$^*$ signals extending further downstream. For the methane-hydrogen mixture, the mean OH$^*$ chemiluminescence signal occupies a larger flame area with relatively more brighter distribution compared to the CH$^*$ chemiluminescence signal, as observed in Fig. \ref{fig_Mean_Flame_OH_CH} (c, d) and (g, h). This is attributed to the hydrogen addition, which has a higher reaction rate with oxygen, leading to increased production of intermediate OH radicals \cite{schefer2003hydrogen,kim2009hydrogen}. 

The flame length is intimately related to the mean flow velocity field, reactant diffusivity and reactive properties of the fuel. For instance, hydrogen exhibits a diffusivity approximately three times greater than that of methane, and its laminar flame speed under stoichiometric conditions is roughly six times higher \cite{dunn2011lean,kobayashi2019science}. According to the theoretical estimation of flame thickness, which scales as $\frac{D_T}{S_L}$ (where $D_T$ is thermal diffusivity and $S_L$ is the laminar flame speed), hydrogen’s higher diffusivity and flame speed result in significantly thinner flames. In contrast, methane’s lower diffusivity and slower flame speed yield broader flames. In addition, the flame length, which also scales with the ratio of exit swirler diameter times the bulk velocity to flame speed ($\frac{D_s U_y}{S_L}$), tends to be shorter in hydrogen-containing flames due to higher flame speed \cite{ho2024direct}. This behavior is clearly observed in the average OH* chemiluminescence images, where methane-hydrogen mixtures produce shorter, more compact flames compared to the elongated structures seen in pure methane combustion as shown in figure \ref{fig_Mean_Flame_OH_CH}.

\begin{figure}[h!]
    \centering
    \hspace*{-5mm}
    \includegraphics[width=0.51\textwidth]{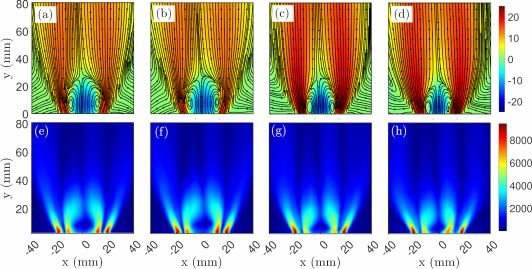}
    \caption{(Top) Mean flow axial velocity field ($\bar{v}_y$ in m/s) with streamlines overlaid and (bottom) strain rate (in 1/s) corresponding to (a,e) CH$_4$: LA, (b,f) CH$_4$: HA, (c,g) CH$_4$-H$_2$: LA, and (d,h) CH$_4$-H$_2$: HA, respectively. }
    \label{fig_PIV_Mean_vel}
\end{figure}

When comparing the chemiluminescence distributions between pure methane and methane-hydrogen combustion, it is observed that pure methane flames exhibit a larger chemiluminescence intensity region. This suggests that the methane flame sustains a broader luminous reaction zone compared to the methane-hydrogen flame. This is attributed to the increased extinction strain rate associated with the hydrogen-methane reactive mixture \cite{shanbhogue2016flame,taamallah2016turbulent}. In addition, this observation is consistent with the axial velocity vector field from Fig. \ref{fig_PIV_Mean_vel}, where the inner recirculation zone (IRZ) in methane-hydrogen combustion was smaller than in pure methane combustion and discussed further in the following paragraphs. The reduced IRZ along with increased extinction strain rate associated with methane-hydrogen flames leads to a more compact reaction region, which aligns with the reduced flame intensity area observed in Fig. \ref{fig_Mean_Flame_OH_CH}. This clearly shows that the flame surface is intimately related to the mean flow velocity field, reactant diffusivity, and reactive properties of the fuel. In both methane and methane-hydrogen fuel combustion, the lower amplitude (LA) OH$^*$ and CH$^*$ chemiluminescence signals exhibit a smaller flame area compared to the higher amplitude (HA) cases. This trend is evident when comparing Fig. \ref{fig_Mean_Flame_OH_CH} (a, e) and Fig. \ref{fig_Mean_Flame_OH_CH} (b, f) for methane, as well as Fig. \ref{fig_Mean_Flame_OH_CH} (c, g) and Fig. \ref{fig_Mean_Flame_OH_CH} (d, h) for methane-hydrogen flames. 

Figure \ref{fig_PIV_Mean_vel} (a-d) illustrates the average axial velocity field overlaid with streamlines for two different combustion conditions: LA and HA cases for both methane and methane-hydrogen mixtures. It is important to note that for the methane cases (refer Fig. \ref{fig_PIV_Mean_vel}(a,b)), the difference between LA and HA conditions is minimal, as the air mass flow rate is nearly the same in both scenarios, with the fuel flow rate held constant. However, in the methane-hydrogen mixture case (refer Fig. \ref{fig_PIV_Mean_vel}(c,d)), a broader range of air mass flow rates is required due to hydrogen’s wider flammability limit. As a result, the gap between LA and HA operating conditions is significantly larger in the methane-hydrogen mixture than in the methane.

For methane, the average velocity field does not show a substantial difference between LA and HA combustion conditions. The maximum velocity and the size of the central recirculation bubble remain almost unchanged, indicating that the transition between LA and HA combustion in methane is not strongly influenced by variations in flow structure as shown in Figs. \ref{fig_PIV_Mean_vel} (a) and (b). In contrast, for the methane-hydrogen mixture, the flame remains unstable over a wider range of operating conditions, leading to noticeable changes in the velocity field as the air mass flow rate increases as shown in Figs. \ref{fig_PIV_Mean_vel} (c) and (d). Specifically, both the velocity magnitude and the size of the recirculation bubble increase with flow rates. The maximum axial velocity for the LA case, i.e., 450 SLPM, is 25.2 m/s, while for the HA case, i.e., 390 SLPM, it is 23.3 m/s. Regarding the recirculation bubble size, the height is measured up to the stagnation point adjacent to the recirculation zone downstream. The LA case exhibits a recirculation bubble height of 26.8 mm, while the HA case shows a smaller height of 23.7 mm.

When comparing the methane and methane-hydrogen cases, an important observation to note is that the recirculation bubble size is smaller for the methane-hydrogen mixture, despite the fact that the air mass flow rates in the CH$_4$-H$_2$ cases are higher than in the pure methane cases. This phenomenon is a direct result of hydrogen’s greater reactivity and higher flame speed compared to methane. The increased reactivity of hydrogen enhances combustion intensity, leading to faster fuel consumption and a reduction in the size of the inner recirculation zone caused by flow expansion in the axial direction \cite{MANSOURI2016}.

Figure \ref{fig_PIV_Mean_vel}(e–h) presents the strain rate fields derived from the corresponding mean velocity fields for both pure methane and methane–hydrogen mixture cases following \cite{shanbhogue2016flame}. In the methane flame, the strain rate is notably higher near the inner recirculation zone and extends further along the axial direction compared to the methane–hydrogen mixture. Although the mean velocity is lower in the methane case, the strain rate field is influenced more significantly by spatial gradients in velocity than by the absolute velocity magnitude.

\subsection{Flame dynamics: Methane HA case \label{CH4_HA_flame_DMD}}

\begin{figure}[h!]
    \centering
    \includegraphics[width=0.49\textwidth]{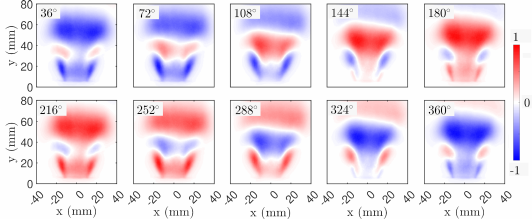}
    \caption{DMD mode of CH$_4$: HA (Fig. \ref{fig_DMD_mode_flame}c) over different phase angles illustrating the ITA mode heat release rate evolution.} 
    \label{fig_DMD_flame_CH4_US}
    \vspace{-3mm}
\end{figure}

Figure \ref{fig_DMD_flame_CH4_US} depict the phase evolution of fundamental thermoacoustic mode for pure methane HA case. The phase evolution of the dominant mode is illustrated using ten equally spaced angles. At a phase angle of 36$^\circ$, the maximum in heat release marked by the red color located downstream of the stagnation point of IRZ. As the phase progresses to 72 and 144 degrees, the maximum in HRR region experiences a downward shift toward the swirler through inner shear layer, influenced by the inner recirculation zone. Simultaneously, the fresh fuel-air mixture enters the reaction zone and subsequently, the maximum in HRR spreading to the inner shear layer close to the flare. At 144 degrees, the maximum in HRR region encompasses nearly the entire recirculation zone also strengthening at the inner shear layer of the flame. This OH$^*$ maximum gradually expands and simultaneously convects downstream, for instance, refer to phase evolutions of $144^\circ-288^\circ$.  During this upstream maximum OH$^*$ intensity propagation process, the spatial minimum in OH$^*$ in inner shear layer convected downstream and further its spatial spread reaches the minimum at 216$^\circ$. This minimum in spatial values subsequently spreads out similar to the maximum in OH$^*$ intensity as discussed above in phase angles of $252^\circ-36^\circ$. It is interesting to note that a sudden increase in maximum and minimum heat release occurs in methane cases caused by a flame front propagation from the tip of the ISL to the IRZ and their convection downstream. In contrary, a gradual shedding with continuous decay behaviour observed in methane-hydrogen case as discussed in section \ref{sec_Flm_DMD}.

\subsection{Flow dynamics: Methane HA case \label{CH4_HA_flow_DMD}}

\begin{figure}[h!]
    \centering
    \hspace{-5 mm}
    \includegraphics[trim={0 10mm 0 15mm},width=0.49\textwidth]{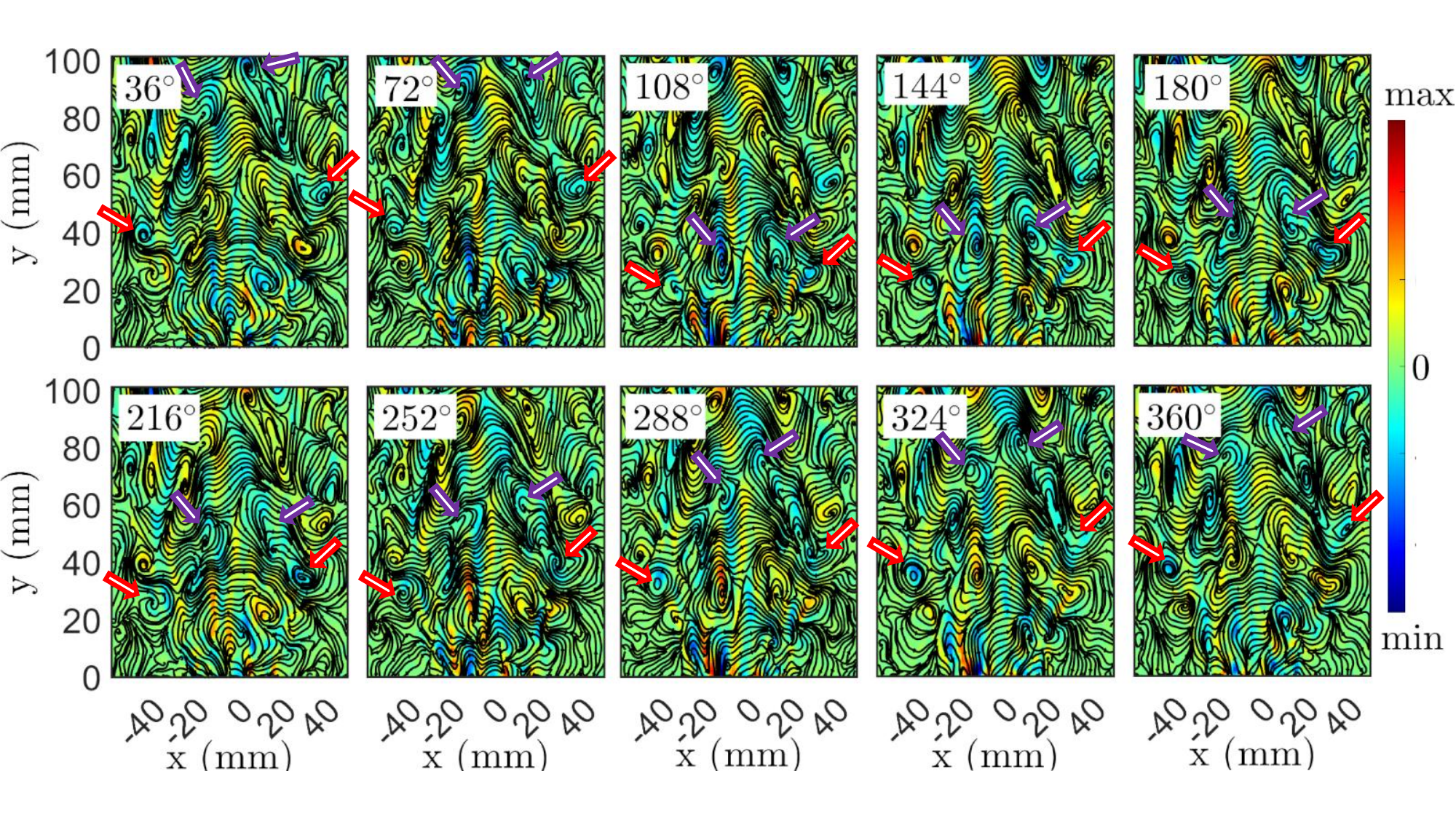}
    \caption{Vorticity field (in 1/s) DMD mode over different phase angles associated with CH$_4$:HA shown in Fig. \ref{fig_DMD_mode_PIV}(b). The purple and red arrow indicated the inner and outer vortex shedding, respectively.}
    \label{fig_DMD_PIV_CH4_US}
    \vspace{-3mm}
\end{figure}

Figure \ref{fig_DMD_PIV_CH4_US} depicts the phase evolution of the dominant thermoacoustic mode for the pure methane case. At a phase angle of 108$^\circ$, distinct vortex structures emerge within the inner and outer shear layers of the recirculation zone. These structures are categorized into two groups: the purple arrow represents vortex pairs convected within the inner shear layer of the recirculation bubble, and the red arrow indicates vortices convected from outer shear layer of the recirculation bubble. As the phase angle progresses, the inner recirculation vortices exhibit a rapid downstream convection compared to the outer recirculation vortices, as indicated by the displacement of the structures marked by the purple and red arrows. This behaviour highlights the dominant influence of the primary swirler on the flow dynamics, as it governs 60\% of the total mass flow rate, whereas the secondary swirler contributes only 40\%. The stronger impact of the primary swirler results in the inner shear layer vortices being convected more rapidly, while the outer recirculation vortex pairs remain relatively stationary for a longer duration. It is interesting to point out that the inner and outer vortex pairs do not interact with each other and convect at different velocities.

\bibliographystyle{cnf-num}
\bibliography{IJHE_2025}

\end{document}